\documentclass[english,aps, twocolumn,amsmath,amssym]{revtex4-1}
\usepackage[T1]{fontenc}
\usepackage[latin9]{inputenc}
\usepackage{textcomp}
\usepackage{amstext}
\usepackage{amssymb}
\usepackage{graphicx}
\usepackage{esint}

\makeatletter

\@ifundefined{textcolor}{}
{%
 \definecolor{BLACK}{gray}{0}
 \definecolor{WHITE}{gray}{1}
 \definecolor{RED}{rgb}{1,0,0}
 \definecolor{GREEN}{rgb}{0,1,0}
 \definecolor{BLUE}{rgb}{0,0,1}
 \definecolor{CYAN}{cmyk}{1,0,0,0}
 \definecolor{MAGENTA}{cmyk}{0,1,0,0}
 \definecolor{YELLOW}{cmyk}{0,0,1,0}
}

\usepackage{units}

\makeatother

\usepackage{babel}
\begin{document}

\title{Vibronic coupling explains the ultrafast carotenoid-to-bacteriochlorophyll
energy transfer in natural and artificial light harvesters}

\author{V\'{a}clav Perl\'{i}k,\textsuperscript{1} Joachim Seibt,\textsuperscript{1}
Laura J. Cranston,\textsuperscript{2} Richard J. Cogdell,\textsuperscript{2}
Craig N. Lincoln,\textsuperscript{3} Janne Savolainen,\textsuperscript{4}
Franti\v{s}ek \v{S}anda,\textsuperscript{1} Tom\'{a}\v{s} Man\v{c}al,\textsuperscript{1}
and J\"{u}rgen Hauer\textsuperscript{3}}

\email[a)To whom correspondence should be addressed: ]{juergen.hauer@tuwien.ac.at}

\affiliation{\textsuperscript{1}Institute of Physics, Faculty of Mathematics
and Physics, Charles University in Prague, Ke Karlovu 5, Prague 121
16, Czech Republic}

\affiliation{\textsuperscript{2}Institute of Molecular Cell and System Biology,
College of Medical, Veterinary and Life Sciences, University of Glasgow,
Glasgow Biomedical Research Centre, 120 University Place, Glasgow
G12 8TA, Scotland}

\affiliation{\textsuperscript{3}Photonics Institute, Vienna University of Technology,
Gusshausstrasse 27, 1040 Vienna, Austria}

\affiliation{\textsuperscript{4}RuhrUniversity Bochum, 44780 Bochum, Germany}

\date{\today}
\begin{abstract}
The initial energy transfer in photosynthesis occurs between the light-harvesting
pigments and on ultrafast timescales. We analyze the carotenoid to
bacteriochlorophyll energy transfer in LH2 \textit{Marichromatium
purpuratum} as well as in an artificial light-harvesting dyad system
by using transient grating and two-dimensional electronic spectroscopy
with 10\,fs time resolution. We find that F\"{o}rster-type models reproduce
the experimentally observed 60\,fs transfer times, but overestimate
coupling constants, which leads to a disagreement with both linear
absorption and electronic 2D-spectra. We show that a vibronic model,
which treats carotenoid vibrations on both electronic ground \textit{and}
excited state as part of the system\textquoteright s Hamiltonian,
reproduces all measured quantities. Importantly, the vibronic model
presented here can explain the fast energy transfer rates with only
moderate coupling constants, which are in agreement with structure
based calculations. Counterintuitively, the vibrational levels on
the carotenoid electronic ground state play a central role in the
excited state population transfer to bacteriochlorophyll as the resonance
between the donor-acceptor energy gap and vibrational ground state
energies is the physical basis of the ultrafast energy transfer rates
in these systems.
\end{abstract}


\keywords{Insert suggested keywords here --- APS authors don't need to do this.}

\maketitle

\section{Introduction\label{sec:Intro}}

Modern researchers have been fascinated by how efficiently solar photons
are converted into chemical energy in photosynthesis. Initially, photons
are absorbed in so-called light-harvesting complexes (LHCs). The following
steps of the cascading energy transfer within LHCs are remarkably
efficient: at low light intensities, 9 out of 10 absorbed photons
create a charge separated state in the reaction center as the basis
for further energy conversion. \cite{Renger2001,Lokstein2013} A thorough
understanding of this process and its technological utilization will
be a major contribution to a sustainable energy concept. Accordingly.
this hope has motivated numerous studies, with the aim of finding
and characterizing a bio-inspired, artificial light harvesting systems.
\cite{Sundstrom2011} Despite the efforts, artificial systems remain
still less efficient and less stable than their natural counterparts.
It is generally accpeted that a detailed picture of the involved electronic
energy levels as well as the energy dissipation pathways in both artificial
and natural light harvesters is crucial in the realization of an efficient
and durable bio-mimetic photosystem.

While there is great structural diversity and flexibility in photosynthetic
LHCs, \cite{Blankenship2014} essentially only two molecular species
serve as pigments, namely carotenoids and (bacterio)chlorophylls,
(B)Chls. Carotenoids and (B)Chls are fundamentally different in structure
and complementing in physiological function. Carotenoids are linear
molecules with varying endgroups. Their optical properties are defined
by the $\pi$-conjugated electronic states, extended along the polyene
backbone as depicted in Fig.~\ref{sec:Figure absorption}. The earliest
suggested energy flow models used a three-level system with ground
state $S_{0}$ and excited states $S_{2}$ and $S_{1}$ \cite{Polivka2004}.
The linear absorption spectra of carotenoids stem from transitions
between the electronic ground state $S_{0}$ and the ``bright''
electronic excited state $S_{2}$, whereas the lowest-lying excited
state $S_{1}$ is ``dark'', with the $S_{0}\rightarrow S_{1}$ transition
being one-photon forbidden. The properties of $S_{1}$ are observed
either by two-photon absorption from $S_{0}$ or by using nonlinear
spectroscopy, e.g. transient absorption (TA), where the ESA $S_{1}\rightarrow S_{n}$
is characteristically strong. \cite{Polivka2009a} The population
flow rate from $S_{2}$ and $S_{1}$ strongly depends on the carotenoid's
chain length, with typical transfer times of less than 200\,fs for
solvated carotenoids (\textit{in vitro}).Besides quenching of harmful
long-lived triplet states in (B)Chls and structural support of LHCs,
the efficient transfer of the excitation energy from carotenoid to
the lower lying (B)Chl states is the main functional role of carotenoids.
The most efficient transfer route occurs from carotenoid's $S_{2}$
state to the energetically closest (B)Chl-state, namely $Q_{x}$.
Transfer from $S_{1}$ to $Q_{y}$ is found in several antenna complexes
and occurs on a longer timescale than $S_{2}\rightarrow Q_{x}$.\cite{Polivka2004}
\textit{In vivo}, i.e. as part of a LHC, the lifetime of $S_{2}$
decreases dramatically. \cite{Polivka2010} For example, in LHCII,
the main antenna system of higher plants and algae, the lifetime of
the $S_{2}$ state of the involved carotenoids is \ensuremath{\sim}120
fs \textit{in vitro} and only \ensuremath{\sim}26 fs \textit{in vivo}
as measured by fluorescence upconversion, which makes it one of the
shortest in naturally occurring systems. \cite{Knox1999} The reason
for this dramatic speed-up in carotenoid lifetime is the interaction
with the closely neighboring (B)Chls. Chemically, the chromophore
of (B)Chls is a nitrogen containing tetrapyrrole-macrocycle with one
or two reduced pyrrole rings. According to the four orbital picture
of porphyrins, \cite{Gouterman1959} the absorption spectrum of (B)Chls
is described by at least two transitions in the so-called Soret band
($B_{x}$ and $B_{y}$ for HOMO - LUMO+1 and HOMO-1 - LUMO+1, respectively)
and two $Q$-band transitions (HOMO - LUMO for $Q_{y}$ and HOMO-1
- LUMO for $Q_{x}$). Depending on molecular structure, the $B$-band
is found in the blue to near-UV spectral region. The $Q$-bands stem
from transitions with orthogonal transition dipole moments and are
energetically degenerate for planar and symmetric porphyrins such
as metal-centered phthalocyanines.\cite{Nemykin2007,Mancal2012} (B)Chls
show a reduced symmetry in comparison, causing a split of the $Q$-band.
For BChl \textit{a} in organic solvents, the $Q_{x}$-band peaks near
600\,nm, while the maximum of $Q_{y}$ is found just below 800\,nm.
Despite their structural differences, carotenoids and (B)Chls fulfill
complementary tasks in photosynthetic antenna complexes. \textit{In
vivo}, excitonic coupling and the polar protein environment shift
the $Q_{y}$ -band of (B)Chls to the red, thus covering the red edge
of the visible solar spectrum. 

In this contribution we investigate an artificial dyad that mimics
\textit{Rhodopseudomonas acidophila} (LH2 \textit{Rps. ac.}), and
compare it to a naturally abundant LH2 of \textit{Marichromatium purpuratum}
(LH2 \textit{M. pur.}, formerly known as \textit{Chromatium purpuratum}).
The main difference between the artificial dyad and LH2 \textit{M.
pur.} is the smaller energy difference between $S_{2}$ and $Q_{x}$
in the latter. Based on this comparison, along with results reported
in the literature for LH2 \textit{Rps. ac.}, we develop a novel energy
transfer model for $S_{2}\rightarrow Q_{x}$ energy transfer. We show
that a transfer process mediated by vibronic resonances between $S_{2}$
and $Q_{x}$ presents a mechanistic alternative to conventional schemes
incorporating only electronic coupling.

\section{Materials and Methods\label{sec:Methods}}

\subsection{Sample preparation}

The dyad-sample was provided by the group around Ana Moore and synthesized
according to previously published procedures. \cite{Macpherson2002,Savolainen2008}
The dyad and its constituents were dissolved in spectroscopic grade
toluene.

Cells of \textit{Marichromatium purpuratum} strain BN5500 (also designated
DSM1591 or 984) were grown anaerobically in the light, harvested by
centrifugation and suspended in 20 mM Tris\textendash HCl pH 8.0.
Then, upon the addition of DNase and MgCl$_{2}$, the cells were broken
by passage through a French press. The photosynthetic membranes were
collected from the broken-cell mixture by centrifugation and resuspended
in the Tris\textendash HCl solution to an optical density (OD) of
0.6 at 830 nm. The membranes were then solubilized by the addition
of 1\%(v/v) of the detergent N, N-dimethyldodecylamine-N-oxide (LDAO).
These solubilized membranes were subjected to sucrose density centrifugation
with a step gradient (as described previously by Brotosudarmo \textit{et
al.},\cite{Brotosudarmo2011}) in order to fractionate the sample
into LH2 and RC\textendash LH1 complexes. The LH2 fraction was then
further purified by ion-exchange chromatography using diethylaminoethyl
cellulose (DE52, Whatman). The bound protein was washed with 20 mM
Tris\textendash HCl pH 9.0 containing 0.15\% dodecyl maltoside (DM)
to exchange the LDAO detergent and then eluted by adding increasing
concentrations of NaCl also containing 0.15\% DM. Additional purification
was achieved by gel-filtration chromatography using a Superdex S-200
column (XK 16/100, GE Healthcare) which had been equilibrated overnight
with 20 mM Tris\textendash HCl pH 9.0 plus 0.15\% DM.

\subsection{Time resolved spectroscopy}

The two spectroscopic techniques employed in this article, transient
grating (TG) and two dimensional electronic spectroscopy (2D-ES) are
both four wave mixing techniques. In four wave mixing, three excitation
pulses interact with the sample to create a non-linear polarization
of third order in the field, which acts as a source term for an emerging
signal. In the experiment employed here, all three excitation fields
emerge along individual wavevectors, resulting in a non-linear signal
direction which is different from any of the incoming beams. This
allows for a virtually background free signal. Details of the experimental
layout are given elsewhere. \cite{Milota2013,Christensson2011} Briefly,
excitation pulses tunable throughout the visible spectral range are
provided by a home-built non collinear optical parametric amplifier
(NOPA), \cite{Piel2006} pumped by a regenerative titanium-sapphire
amplifier system (RegA 9050, Coherent Inc.) at 200 kHz repetition
rate. Pulse spectra were chosen to overlap with the investigated sample's
absorption spectrum (see Fig.\,\ref{sec:Figure absorption}) and
compressed down to a temporal FWHM of sub-11 fs in each case, determined
by intensity autocorrelation. The pulses were attenuated by a neutral
density filter to yield 8.5\,nJ per excitation pulse at the sample.
This corresponds to a fluence of less than $3.0\times10^{14}$ photons/cm$^{2}$
per pulse or 0.5\% of excited molecules in the focal volume. The four
wave mixing experiment used for both TG and 2D-ES relies on a passively
phase stabilized setup with a transmission grating \cite{Cowan2004,Brixner2004b}
and has a temporal resolution of 5.3\,as for $t_{1}$, i.e. the delay
between the first two pulses and 0.67\,fs for $t_{2}$, the delay
between second and third excitation pulse. A detailed description
can be found in Milota \textit{et al.}\cite{Milota2009a} For TG measurements,
$t_{1}$, was kept at 0\,fs while $t_{2}$ was scanned. The emerging
signal was spectrally resolved in $\omega_{3}$ by a grating-based
spectrograph and recorded with a CCD camera. At a given delay time,
spectra were not recorded on a shot-to-shot basis, but averaged over
approximately $10^{5}$ shots per spectrum. Sample handling was accomplished
by a wire-guided drop jet \cite{Tauber2003} with a flow rate of 20
ml/min and a film thickness of approx. 200\,\textmu m. The fact that
the jet operates without the use of cell windows allows us to interpret
2D spectra and TG-signals even during pulse overlap ($t_{1}=t_{2}=0$).
All measurements were performed under ambient temperatures (295 K).

\subsection{Modeling}

\subsubsection{Modeling of 2D-spectra}

To facilitate the interpretation of the measured 2D-spectra, results
of model calculations were included in the analysis. It turned out
that simulations including only carotenoid allowed most features of
the measured 2D spectra to be explained. Therefore the model was based
on the properties of the carotenoid component, whereas transfer from
carotenoid to BChl only entered in terms of a contribution to lifetime
broadening. With regards to electronic transitions it was assumed
that the spectral profile of the pulses used in the experiment allows
for resonant transitions between $S_0$ and $S_2$, and from $S_1$
to $S_n$ subsequent to population transfer from $S_2$ to $S_1$
. To account for vibrational effects in excitation processes involving
the electronic states $k$ and $l$, two modes were included as underdamped
oscillators with reorganization energies $\lambda_{UO,i}, \; i \in \{ 1,2 \}$
and vibrational frequencies $\omega_{UO,i}, \; i \in \{ 1,2 \}$ in
terms of line shape functions \cite{Mukamel95}

$$g_{UO, kl,i}(t)=\frac{\lambda_{UO,kl,i}}{\omega_{UO,i}}$$
$$\times\Big( \coth \left( \frac{\hbar \omega_{UO,i}}{k_B T} \right) \left( 1-\cos(\omega_{UO,i} t) \right)$$
\begin{equation} \label{eq:line_shape_function_underdamped_oscillator} 
+ i \sin(\omega_{UO,i} t) -i \omega_{UO,i} t \Big). 
\end{equation}The reorganization energies are connected to the Huang-Rhys factors
$\chi_{UO,kl,i}$ via $\lambda_{UO,kl,i}=\chi_{UO,kl,i} \omega_{UO,i}$.
Damping of the vibrations, which is known to appear in $S_1$, was
included by calculating the line shape function of the respective
underdamped oscillator via double integration of a correlation function
\cite{Mukamel95}, where the real part of the correlation function
was approximated by including terms up to a selected order $n$ of
a series characterized by the Matsubara frequencies $\nu_n=\frac{2 \pi k_B T n}{\hbar}$.

Furthermore, during the evolution in electronic states $k$ and $l$
the influence of the bath was described by two Brownian oscillator
spectral density components with damping constant $\Lambda_{k l,i}, \; i \in \{ 1,2 \}$
and reorganization energy $\lambda_{BO,k l,i}, \; i \in \{ 1,2 \}$.
From the respective spectral density components \cite{Mukamel95}

\begin{equation} \label{eq:spectral_density_Brownian_oscillator} J_{BO, k l,i}(\omega)=2 \lambda_{BO,k l,i} \frac{\omega \Lambda_{k l,i}}{\omega^2+\Lambda_{k l,i}^2} \end{equation}line
shape functions $g_{BO, k l,i}(t)$ of the bath can be calculated
by using the general formula for an arbitrary spectral density $J(\omega)$
, which reads

\begin{equation} \label{eq:line_shape_function} 
\begin{split} 
g(t)&=\frac{1}{2 \pi} \int^{\infty}_{-\infty} d \omega \frac{1-\cos(\omega t)}{\omega^2} \coth \left( \frac{\omega}{2 k_B T} \right) J(\omega) \\ &+\frac{i}{2 \pi} \int^{\infty}_{-\infty} d \omega \frac{\sin(\omega t)-\omega t}{\omega^2} J(\omega). 
\end{split} 
\end{equation}The sum of all line shape function components was included in the
response functions for the calculation of 2D-spectra. The response
functions contain the transition frequencies $\omega_{k l}$ between
the electronic states $k$ and $l$ and rate constants of transfer
processes. Besides the rate $k_{S_2 \to S_1}$ for transfer from $S_2$
to $S_1$ also a transfer rate $k_{S_2 \to Q_x}$ from $S_2$ to an
electronic state $Q_x$ of the Chlorophyll component of the LH2 complex
plays a role in lifetime broadening effects.

The rate of lifetime broadening in $S_2$ consists of the sum $\frac{1}{2}(k_{S_2 \to S_1}+k_{S_2 \to Q_x})$
and is denoted as $\Gamma_{S_2}$. For lifetime broadening in $S_1$
a rate $\Gamma_{S_1}$ was introduced. The response functions can
be distinguished by their rephasing or nonrephasing properties and
assigned to excitation processes of stimulated emission (SE), ground
state bleaching (GSB) or excited state absorption (ESA) type, where
in the case of SE and GSB only the electronic transition dipole moment
$\mu_{S_0 S_2}$ enters, while for ESA components with population
transfer the transition dipole moment $\mu_{S_1 S_n}$ also appears.

The response functions were formulated in analogy to \cite{SePu14_JCP_114106}
and \cite{MaDoPs14_CJC_135} and are given in the Appendix \ref{sec:Response-functions}.
The sum of all rephasing and nonrephasing response function components
yields the third-order response $S_R^{(3)}(\tau_3,\tau_2,\tau_1)$
and $S_{NR}^{(3)}(\tau_3,\tau_2,\tau_1)$, respectively, which reflect
the excitation dynamics in the limit of infinitely short pulses. The
influence of the finite pulse width pulses was taken into account
in analogy to Ref. \cite{BrMaSt04_4221_}, i.e. by convolution of
the calculated signal with the pulses. By integration over the time
intervals $\tau_1$, $\tau_2$ and $\tau_3$ between the electronic
transitions, the resulting third-order polarization is transformed
into a dependence on the pulse delays $t_1$, $t_2$ and $t_3$. The
rephasing part of the 2D-spectrum is obtained via 

$$\sigma_{2D,R}(\omega_1,t_2,\omega_3)=\int^{\infty}_{0} d t_1  \int^{\infty}_{0} d t_3 \exp(-i \omega_{1} t_1) $$
\begin{equation} \label{eq:rephasing_contribution}
\times \exp(i \omega_{3} t_3) P_{R}^{(3)}(t_1,t_2,t_3), 
\end{equation}the nonrephasing part results from

$$\sigma_{2D,NR}(\omega_1{},t_2,\omega_3)= \int^{\infty}_{0} d t_1  \int^{\infty}_{0} d t_3 \exp(i \omega_{1} t_1) $$
\begin{equation} \label{eq:rephasing_contribution} 
\times \exp(i \omega_{3} t_3) P_{NR}^{(3)}(t_1,t_2,t_3). 
\end{equation}

\subsubsection{Calculation of Energy Transfer Rate in a Vibronic Model\label{sub:Calculation-of-Transfer}}

To simulate the energy transfer rate from the carotenoid $S_{2}$
state to the $Q_{x}$ of the BChl in LH2 and the dyad we employ a
hetero-dimer model. This approach is natural for the dyad showing
now signs of aggregation between the chromophores. For LH2 the monomeric
treatment of $Q_{x}$ and $S_{2}$ is also justified, given that there
are no signs of excitonic splitting for these bands. Moreover the
carotenoid predominantly interacts with one of its neighboring B850
BChls with the next nearest neighbor coupling showing only half the
interaction strength. \cite{Tretiak2000}

To model the carotenoid molecule we use an effective two state model
which involves two electronic states ($S_{0}$ and $S_{2}$) and one
effective vibrational mode which replaces the two fast vibrational
modes with frequencies $\Omega_{1}\approx1150$ cm$^{-1}$ and $\Omega_{2}\approx1500$
cm$^{-1}$ known to be present the carotenoid and it simulates the
progression of states $|0\rangle_{1500}|0\rangle_{1150}$, $|0\rangle_{1500}|1\rangle_{1150}$,
$|1\rangle_{1500}|0\rangle_{1150}$, $|0\rangle_{1500}|2\rangle_{1150},|1\rangle_{1500}|1\rangle_{1150},|2\rangle_{1500}|0\rangle_{1150},\dots$
which originates from the two modes. We can group these levels according
to vibrational energies equal approximately average of two mode values
to $0$, $\approx1325$ cm$^{-1}$, $\approx2650$ cm$^{-1}$, etc.
(counted from the zero point energy). Because the spectral lines corresponding
to the transitions to the states listed above are broadened, they
act effectively as one mode with a roughly average frequency. This
carotenoid mode (referred to as primary mode further on in this paper)
is coupled to a bath of harmonic oscillators \cite{Perlik2014,Landau1936,Sanda2002}
(a secondary modes or secondary bath) which provides damping and dephasing.
The coupling between the effective mode and the carotenoid electronic
transition provides the optical dephasing leading to the broadening
of the absorption spectra. In order to calculate correctly the energy
transfer rate, we parametrize the fast mode and its coupling to the
bath of harmonic oscillators in such a way that it reproduces absorption
spectra. This effective model of the bath enables us to limit the
bath description to one harmonic mode whose dynamics is treated explicitly
by a master equation according to \cite{Perlik2014}. In this way
the mode effectively provides a correct bath spectral density for
the energy transfer between the carotenoid and BChl while allowing
for a non-perturbative treatment. Fit of the absorption spectra of
the carotenoid in the dyad yields the carotenoid effective vibrational
frequency $\omega_{{\rm eff}}=1390$ cm$^{-1}$, HR factor $s_{{\rm eff}}=1.3$,
the reorganization energy of the secondary bath $\lambda_{b}=670$
cm$^{-1}$ and the bath correlation time $\tau_{b}=30$ fs. For the
LH2 \emph{M. pur. }we obtained $\omega_{{\rm eff}}=1453$ cm$^{-1}$,
$s_{{\rm eff}}=1.21$, $\lambda_{b}=1242$ cm$^{-1}$ and $\tau_{b}=54$
fs.

The $Q_{x}$ states of the BChl and purpurin are treated by the same
model. The representative mode on the BChl and purpurin molecules
are chosen so that they fit absorption spectra including the significant
vibrational side bands of the molecules. The side band for the purpurin
can be easily located in the monomeric absorption spectrum given in
Fig. \ref{sec:Figure absorption}. By fitting the purpurin monomeric
absorption spectrum we get the effective mode frequency $\omega_{purp}=1246$
cm$^{-1}$, HR factor $s_{purp}=0.61$, reorganization energy $\lambda_{purp}=600$
cm$^{-1}$ and bath correlation time of $\tau_{purp}=47$ fs. BChl
absorption spectrum exhibits a much weaker vibrational side band and
leads to a smaller HR factor of $s_{BChl}=0.22$. Other parameters
are as follows: $\omega_{BChl}=1284$, $\lambda_{BChl}=1911$ cm$^{-1}$
and $\tau_{BChl}=100$ fs. The effective vibrational modes represent
the overdamped part of the spectral density found in BChls (see e.g.
Ref. \cite{Renger2002}) by the term discussed in Section \ref{sub:Vibronic-Coupling}
and the side band by its first replica. The strength of this replica
is given by the HR factor. 

The parameters obtained by the fit of the monomers are slightly readjusted
in a fit of the aggregate spectra, either the purpurine-carotenoid
dyad or the LH2 complex. Energy transfer rates are then estimated
with no further adjustment of the parameters. To estimate the energy
transfer rate, the population dynamics of the system is calculated
by the master equation from Ref. \cite{Sanda2002} with the initial
population corresponding to the laser pulse excitation of the carotenoid.
The total time dependent population of the $Q_{x}$ state which is
initially zero is calculated and the population transfer time is defined
as the time $\tau_{S_{2}\rightarrow Q_{x}}$ when the population of
$Q_{x}$ state is equal to $1/e$ .

\section{Results\label{sec:Results}}

\subsection{Absorption}

\begin{figure}
\includegraphics{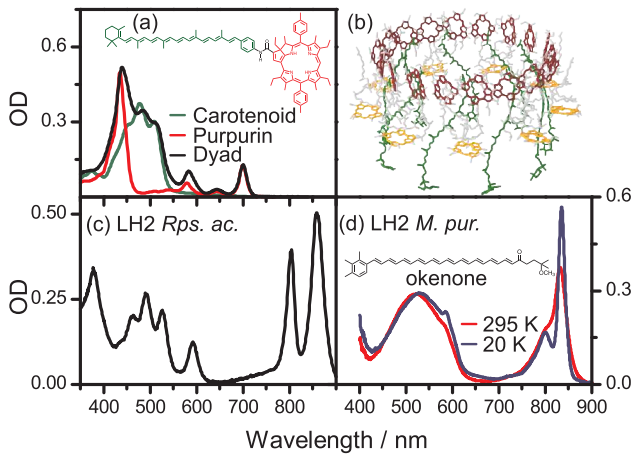}\protect\caption{(a) Absorption spectra and molecular structure of the caroteno-purpurin
dyad (black line) and its constituting carotenoid (donor, green) and
porphyrinic acceptor in red. (b) Structural model and absorption spectrum
(c) of LH2 \textit{Rps. ac. }(d) Absorption spectra at different temperatures
of LH2 \textit{M. pur.}, along with the molecular structure of the
bound carotenoid (okenone). \label{sec:Figure absorption}}
\end{figure}

Fig. \ref{sec:Figure absorption}(a) shows the absorption spectrum
of the dyad (black) and its constituents, i.e. the carotenoid donor
(green), and the purpurin acceptor (red).

Details on the dyad\textquoteright s structure and dynamics are given
in detail elsewhere. \cite{Macpherson2002,Savolainen2008,Savolainen2008a}
It is a donor-acceptor system, consisting of a $\beta$-carotene derivative
(donor, green in the molecular structure in Fig.\ref{sec:Figure absorption}(a)),
and a tetrapyrrole-macrocycle as an acceptor, referred to as purpurin
(red). The dyad-carotenoid and purpurin are linked by an amide bond,
providing structural rigidity but only partial conjugation between
donor and acceptor. The dyad was synthesized to mimic LH2 \textit{Rps.
ac.}, whose structure is depicted in Fig.\,\ref{sec:Figure absorption}(b),
as taken from the Papiz \textit{et al.}\cite{Papiz2003} LH2 \textit{Rps.
ac.} consists of nine pairs of pigment-protein subunits ($\alpha/\beta$-subunits),
forming two concentric cylinders, to which carotenoids and BChl \textit{a}
molecules are bound non-covalently. As can be seen in Fig. \ref{sec:Figure absorption}(b),
the BChls are arranged in two ring-like structures (red and yellow)
with different numbers of BChls (18 for red, 9 for yellow) and ring
diameters, leading to a weakly and strongly interacting set of BChls
(yellow and red, respectively). As a result, the excitonic part of
the absorption spectrum of LH2 \textit{Rps. ac.}, shown in Fig.\,\ref{sec:Figure absorption}(c),
exhibits two peaks, one near the monomeric $Q_{y}$-transition at
800\,nm and a red-shifted peak near 850\,nm. The red and yellow
sets of BChls in Fig. \ref{sec:Figure absorption}(b) are therefore
referred to as B850 and B800, respectively. The carotenoid is in van
der Waals contact (<3.5\,�) with both the B800 and the B850 ring.\cite{McDermott1995,Papiz2003}
The dyad lacks such excitonic bands, as its structure is monomeric
and there are no interacting BChl-moieties. An interesting difference
between the absorption spectrum of the dyad and LH2 \textit{Rps. ac.}
arises in the carotenoid region of the spectrum. The vibronic structure
of the monomeric carotenoid\textquoteright s absorption spectrum,
shown in green in Fig.\,\ref{sec:Figure absorption}(a), is more
pronounced than in the dyad, shown in black. This can be understood
by invoking vibronic coupling between donor and acceptor, as will
be addressed in detail in section \ref{sec:Discussion}. 

The absorption spectrum of LH2 \textit{M. pur. }(Fig.\,\ref{sec:Figure absorption}(d))
exhibits substantially less vibronic modulation in carotenoid region
of the spectrum. This is attributed to the polar endgroup of the bound
carotenoid (okenone), interacting strongly with the polar \textit{in
vivo} protein environment. As the bright $S_{2}$-state carries ionic
character,\cite{Tavan1987} its energy relative to $S_{0}$ is strongly
affected by the interaction with the protein and shifts to the red
in comparison the carotenoid (rhodopin glucoside) of LH2 \textit{Rps.
ac.} The BChl \textit{a} molecules in LH2 \textit{M. pur. }don't show
similar bathochromic shifts, which leads to a greatly enhanced overlap
of the carotenoid- and $Q_{x}$-band in this system. The excitonic
region of the spectrum of LH2 \textit{M. pur. }exhibits a less pronounced
red shift with respect to the monomeric transition as compared to
LH2 \textit{Rps. ac.} Recent preliminary X-ray structure analysis\cite{Cranston2014}
explains this by a reduced number of $\alpha/\beta$-subunits (eight
for LH2 \textit{M. pur. }and nine in the case of LH2 \textit{Rps.
ac.})

\subsection{Transient Grating}

\begin{figure}
\includegraphics{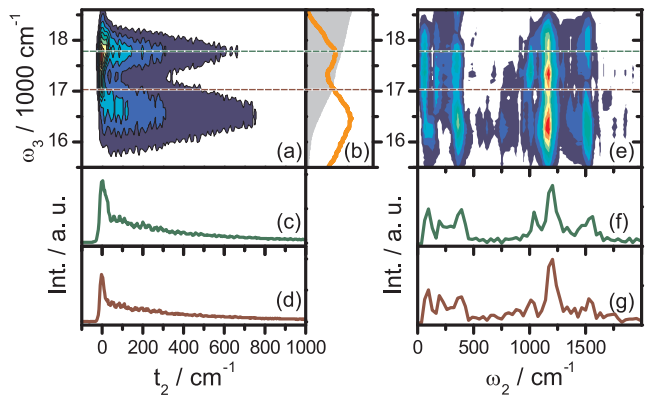}\protect\caption{(a) Detection frequency ($\omega{}_{3}$) resolved transient grating
signal of LH2 \textit{M. pur.} (b) Absorption spectrum in gray in
comparison to the spectrum of the excitation pulse in orange. Temporal
cuts at $S_{2}$-specific detection frequencies specific for $S_{2}$
and $Q_{x}$ in (c) and (d), respectively. (e) Fourier-transform spectra
along $t_{2}$, dispersed along $\omega{}_{3}$. (f) and (g) show
cuts through (e) for $S_{2}$- and $Q_{x}$-specific detection frequencies,
respectively. \label{sec:Figure TG}}
\end{figure}

Fig. \ref{sec:Figure TG}(a) shows the emission frequency ($\omega{}_{3}$)
resolved transient grating (TG) signal of LH2 \textit{M. pur}. The
signal shows a broad frequency range, spanning the entire spectrum
of the excitation pulse, shown as an orange line in Fig. \ref{sec:Figure TG}(b),
in comparison to the absorption spectrum of LH2 \textit{M. pur}. depicted
as a grey area. The double-peaked structure of the excitation spectrum
explains the overall shape of the TG-signal. The dashed lines mark
frequencies characteristic for $S_{2}$ (dashed green line) and $Q_{x}$
(dashed dark red line). Figures \ref{sec:Figure TG}(c) and (d) show
the respective transients. For both detection frequencies, we observe
a fast decaying peak around $t_{2}$\,=\,0\,fs, followed by a slower
mono-exponential decay with a decay constant of 330\,\textpm \,50\,fs
according to least square fit analysis. The signal does not decay
to zero within the 1000\,fs time window of the experiment. The observed
decay behavior is, within error margins, identical for both detection
frequencies. TG probes the sum of the absorptive and dispersive part
of the induced third order signal, while pump-probe selects only the
absorptive part, meaning that the retrieved time constants cannot
be directly compared to pump-probe measurements. \cite{Polli2006,Sugisaki2010}
Vibrational dynamics however will yield similar frequencies between
the two techniques (see. i.e. vibrational dynamics in $\beta$-carotene
probed by pump-probe \cite{Cerullo2001} and by TG \cite{Hauer2006a}).
The vibrational response manifests as an oscillatory signal, superimposed
on a slowly varying background. After subtraction of the latter and
Fourier transformation of the remaining signal along $t_{2}$ for
every detection frequency, the ($\omega_{2}$ , $\omega_{3}$ ) frequency
map in Fig.\,\ref{sec:Figure TG}(e) is retrieved. A cut along a
$S_{2}$-specific detection frequency (green dashed line) yields the
well-known carotenoid frequencies as depicted in Fig.\,\ref{sec:Figure TG}(f)
near 1000\,cm$^{-1}$ (Methyl-rocking motion), 1160\,cm$^{-1}$
(carbon single bond stretching) and 1530\,cm$^{-1}$ (carbon double
bond stretching), and are in agreement with Resonance Raman spectra
measured for okenone. \cite{Fujii1998} The same set of frequencies
is observed when detecting at the $Q_{x}$-band, as shown in Fig.\,\ref{sec:Figure TG}(g).
This is interesting because BChl-specific modes near 1600\,cm$^{-1}$
or 700\,cm$^{-1}$ as know from resonance Raman measurements \cite{Cotton1981}
are not found. The absence of BChl vibrational signatures is readily
explained by vibrational-electronic coupling in BChl, which is small
in comparison to carotenoids.\cite{Christensson2009} 

\subsection{2D electronic spectroscopy}

\begin{figure}
\includegraphics[width=1\columnwidth]{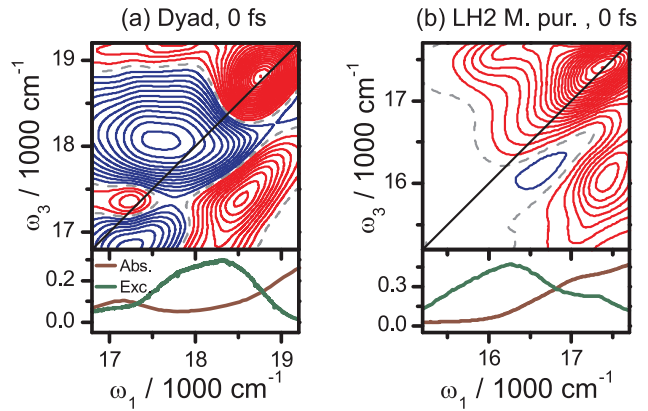}\protect\caption{2D electronic spectra of the dyad (a) and LH2 \textit{M. pur.}, each
at $t_{2}$=\,0\,fs. Positive (negative) signals are drawn a red
(blue) full lines at 5\% steps. Each figure is normalized to its respective
maximum. The dashed grey lines mark the nodal line at zero intensity.
The lower panels in each graph show the overlap between absorption
(dark red line) and excitation spectrum (green line).\label{sec:2D-dyad_vs_LH2}}
\end{figure}

As mentioned in section \ref{sec:Methods}, 2D-ES and TG are related
in the sense that they are both four wave mixing techniques. The difference
lies in the treatment of the first inter-pulse delay $t_{1}$, which
is kept at zero in TG, but scanned and Fourier-transformed ($t_{1}\rightarrow\omega_{1}$)
in 2D-ES. The emerging ($\omega_{1}$, $\omega_{3}$)-2D plots, recorded
at a fixed second inter-pulse delay $t_{2}$, allow for a correlation
of excitation ($\omega_{1}$) and emission frequencies ($\omega_{3}$).
Fig.\,\ref{sec:2D-dyad_vs_LH2} compares electronic 2D spectra at
$t_{2}$\,=\,0\,fs for the dyad and LH2 \textit{M. pur}. 

At $t_{2}$\,=\,0\,fs, electronic population is given no time to
relax to lower lying states, within experimental time resolution.
Both spectra in Fig.\,\ref{sec:2D-dyad_vs_LH2} show strong positive
(red) peaks along the diagonal ($\omega_{1}$=$\omega_{3}$). Comparison
to the absorption spectra, given in the lower panels in Fig.\,\ref{sec:2D-dyad_vs_LH2}(a)
and (b), show that the strong diagonal peaks stem from the carotenoid's
$S_{0}\rightarrow S_{2}$ transition. The dyad's spectrum in Fig.\,\ref{sec:2D-dyad_vs_LH2}(a)
shows two well resolved peaks corresponding to the $Q_{x}$-transition
at approximately 17200\,cm$^{-1}$ and a carotenoid peak near 19000\,cm$^{-1}$.
The off-diagonal or cross peaks ($\omega_{1}\neq$$\omega_{3}$) with
negative amplitude in Fig.\,\ref{sec:2D-dyad_vs_LH2} stem from ESA-pathways.
For carotenoids, ESA from $S_{2}$ in the visible has been observed.
\cite{Christensson2010a,Kosumi2005} The ESA of the dyad's porphyrinic
acceptor was described previously as broad and featureless. While
the combination of donor and acceptor ESA explains the negative features
in the dyad's 2D-spectrum, we note that at $t_{2}$\,=\,0\,fs,
negative signals do not necessarily have to be related to ESA, as
they already occur in a two level system at this waiting time. \cite{Jonas2003}
Positive off-diagonal peaks can stem from vibrational energy levels,\cite{Nemeth2010,Christensson2011,Turner2011a,Butkus2012}
stimulated emission (SE) or coupling between two electronic oscillators.\cite{Jonas2003}
The latter would be the most interesting scenario, as electronic coupling
between donor and acceptor would suggest an excitonic energy level
structure and the associated transfer mechanisms.\cite{Valkunas2013}
Upon visual inspection, the location of the positive cross peaks in
the dyad's 2D signal suggests electronic coupling as a likely explanation.
The stronger of the two positive cross peaks, found below the diagonal
($\omega_{1}>\omega_{3}$), peaks at a value of $\omega_{3}$that
coincides with the absorption frequency of the $Q_{x}$-band. There
is also a corresponding peak above the diagonal ($\omega_{1}<\omega_{3}$),
which peaks blue detuned with respect to $Q_{x}$ and with weaker
intensity than its counterpart below the diagonal. Such differences
in intensity may however be attributed to the ultrafast population
transfer from $S_{2}$ to $Q_{x}$, the onset of which takes place
within the experimental time resolution of 11\,fs. The blue detuned
maximum of the upper cross peak can be attributed to finite pulse
width or effects of chirp in the excitation pulses. \cite{Christensson2010}
Inspection of the 2D signal of LH2 \textit{M. pur }disproves the assumption
of electronic coupling. The major difference between the dyad and
LH2 \textit{M. pur }is the position of the $S_{2}$-transition. While
the dyad's carotenoid exhibits maximum absorption above 19000\,cm$^{-1}$,
the carotenoid in LH2 \textit{M. pur}, okenone, has a carbonyl-containing
endgroup as shown in Fig.\,\ref{sec:Figure absorption}(d), shifting
the 0-0 transition of okenone \textit{in vivo} down to 17570\,cm$^{-1}$,
\cite{Andersson1996} while the energetic position of the $Q_{x}$-band
remains unchanged between dyad and LH2 \textit{M. pur}. Hence, the
energy difference between $S_{2}$ and $Q_{x}$ should be greatly
reduced in LH2 \textit{M. pur}, and the corresponding cross peaks
should be shifted, leading to a large and fairly featureless electronic
2D-signal. In contrast to this expectation, the lower cross peak in
LH2 \textit{M. pur }shows roughly the same energy difference from
the diagonal of approximately 1500\,cm$^{-1}$. This strongly suggest
a vibrational origin, given that the prominent C=C stretching mode
of carotenoids is found at this energy. 

\begin{figure*}
\includegraphics[width=2\columnwidth]{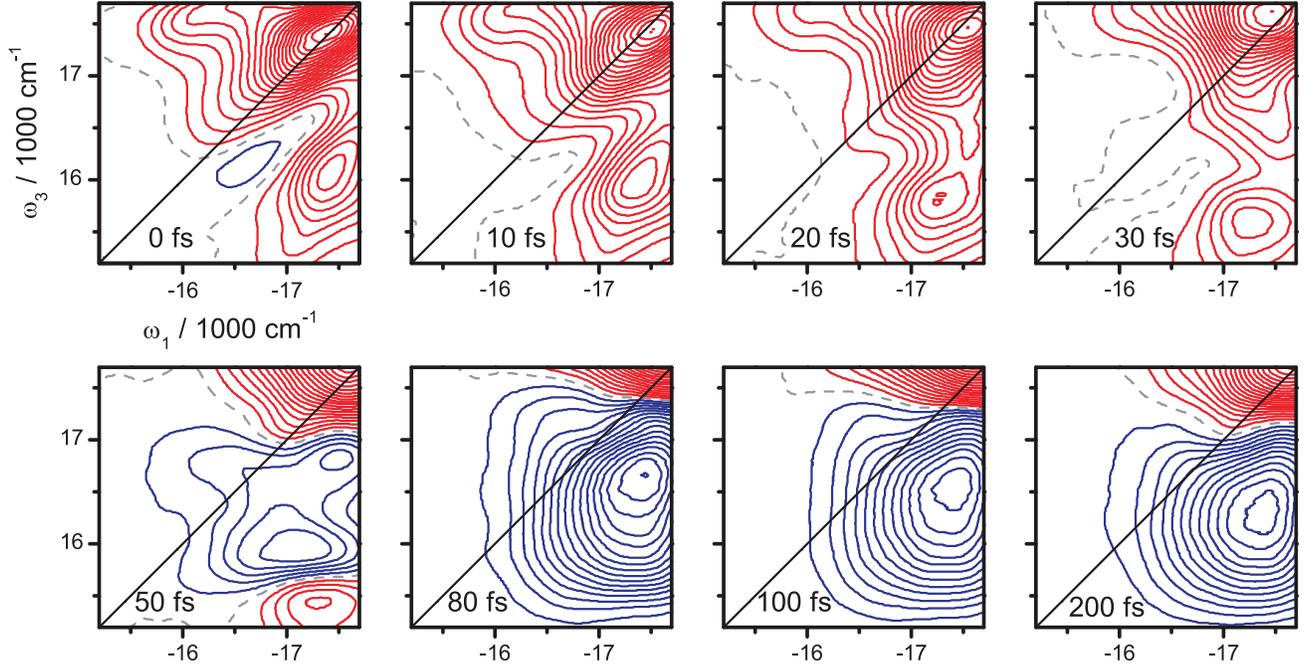}\protect\caption{Electronic 2D spectra of LH2 \textit{M. pur.}, measured at indicated
values of $t_{2}$. Line coloring follows the same conventions as
described in Fig.\,\ref{sec:2D-dyad_vs_LH2} \label{sec:Figure 2D_experimental}.}
\end{figure*}

The peak above the diagonal in LH2 \textit{M. pur }is harder to explain.
The ESA-signal in Fig.\,\ref{sec:2D-dyad_vs_LH2}(b) is much weaker
than for the dyad case, which is readily explained by a different
ESA-transition energy from $S_{2}$ for okenone \textit{in vivo}.
\cite{Polli2006} We can thus observe a weak diagonal peak at $\omega_{1}=\omega_{3}\approx$16480\,cm$^{-1}$,
with an excitation frequency $\omega_{1}$ corresponding to the upper
cross peak. This energetic position excludes coupling to $Q_{x}$
as a possible explanation, peaking at 17035\,cm$^{-1}$ in LH2 \textit{M.
pur}. The physical origin of the upper cross peak in Fig.\,\ref{sec:2D-dyad_vs_LH2}(b)
becomes more obvious when examining the peak shape evolution along
$t_{2}$, as shown in Fig.\,\ref{sec:Figure 2D_experimental}. 

The upper diagonal peak is discernible up to 30\,fs, but drops to
zero by 80\,fs. This behavior is unexpected for an electronic coupling
peak, whose amplitude should only be a function of coupling constant
\textit{J}, but not of waiting time $t_{2}$. The rapid decay of the
upper cross peak rather suggests that it is an artifact, related to
the shape of the excitation pulses. The excitation spectra are clearly
non-Gaussian (see lower panel of Fig.\,\ref{sec:2D-dyad_vs_LH2}(b)),
making limited pulse width or chirp related effects \cite{Christensson2010}
the most likely explanation for the upper diagonal cross peak for
LH2 \textit{M. pur}. 

Another interesting feature is observed at ($\omega_{1}$\,=\,-17500\,cm$^{-1}$,
$\omega_{3}$\,=\,16480\,cm$^{-1}$), i.e. an energetic position
indicative of $S_{2}\rightarrow Q_{x}$ energy transfer pathways.
The peak builds up rapidly within the first 30\,fs, which corresponds
well with the 55\,fs transfer time reported for this process.\cite{Polli2006}
After 50\,fs, negative ESA-signals dominate the spectra. This is
explained by the intramolecular $S_{2}\rightarrow S_{1}$ energy transfer,
occurring at a 95\,fs build up rate as reported previously.\cite{Polli2006}
We support this assignment by calculating the electronic 2D-spectra
only for okenone, the carotenoid of LH2 \textit{M. pur}. Details of
the calculations can be found in section \ref{sec:Methods}C.

\section{Discussion\label{sec:Discussion}}

\subsection{Carotenoid-Chlorophyll Interaction\label{sub:Carotenoid-Chlorophyll-Interaction}}

Spectroscopic experiments reveal the transfer from Car $S_{2}$ and
$S_{1}$ states to Chls and BChls as fast and relatively efficient
despite the rather short $S_{2}$ state life time of the monomeric
carotenoid (at least in comparison with Chls and BChls). Both $Q_{x}$
and $S_{2}$ states are optically allowed, and they can therefore
interact by the resonance coupling mechanism. Because of the proximity
of the two molecules in the LH2 structure, one cannot exclude an electron
exchange (Dexter) mechanism. However, quantum chemical calculations
suggest Coulomb interaction to be dominant here \cite{Nagae1993}.
In this case, theoretical description the carotenoid-BChl interaction
and modeling of the corresponding excitation energy dynamics remain
within the well understood framework of excitonic description of photosynthetic
antennae \cite{VanAmerongen2000,Valkunas2013}. 

In the standard framework, we assume the electronic coupling element
between the two electronic states to be independent of the vibrational
DOF of the molecular system. Electronic excitation is transferred
between the two molecules due to an interplay of the resonance coupling
and the fast fluctuations of the molecules' respective energy gaps.
The energy gap fluctuations are caused by their interaction with the
nuclear degrees of freedom (DOF) which involve both the intra- as
well as intermolecular vibrations. These are considered a thermodynamic
heat bath characterized by certain temperature and spectral density.
Two regimes of the relative coupling strengths are usually distinguished.
The \emph{delocalized regime} in which the system-bath interaction
is weak in comparison with the resonance interaction between the molecules,
and the\emph{ localized regime} in which the system-bath coupling
is considered strong. The terms localized and delocalized refer to
the effective electronic eigenstates of the interacting system. For
the delocalized regime the coupling between the two molecules is strong
enough to sustain correlation between the electronic excitations on
different molecules so that delocalized excitons are formed. In the
localized regime (also referred to as F\"{o}rster regime here), the system-bath
interactions induce energy gap fluctuations which prevent any such
correlation, and the excitations appear to be localized on the molecules.
In fact, even in relatively weakly excitonically coupled systems one
can often detect signs of delocalization (see e.g. Ref. \cite{Cheng2006}),
because the actual apparent eigenstates are always somewhere in between
the two theoretical limits. For photosynthetic aggregates formed by
BChl and Chl molecules, where the protein bath can be to some extent
assumed as unstructured, modern simulation methods such as the Hierarchical
Equations of Motion (HEOM) enable us to determine the excited state
dynamics beyond the two limits mentioned above.\cite{Ishizaki2009}
However, even in the case of aggregates composed of BChl or Chl molecules
only, some clearly visible spectroscopic features can result from
the involvement of pronounced vibrational modes. \cite{Christensson2012,Tiwari2013,Chin2013,Chenu2013} 

In the discussion that follows, we want to argue that the involvement
of the vibrational DOF of the carotenoid molecules is crucial for
the energy transfer dynamics between the $S_{2}$ and the $Q_{x}$
states. We will compare two theoretical approaches to the calculation
of the energy transfer rates, both motivated by the observed features
of the spectra and known spectroscopic behavior of the system. We
will argue that both support the crucial role of the fast vibrational
modes, in particular the carotenoid ground state vibration levels. 

The Hamiltonian of the relevant part of the system (state $Q_{X}$
of the BChl and the $S_{2}$ state of the carotenoid) can be written
in the following way
\[
H=H_{Q_{x}}^{(B)}+H_{S_{2}}^{(B)}+\left(\epsilon_{Q_{x}}+\Delta V_{Q_{x}}(q^{(Q_{x})})\right)|Q_{x}\rangle\langle Q_{x}|
\]
\[
+\left(\epsilon_{S_{2}}+\Delta V_{S_{2}}(q^{(S_{2})})\right)|S_{2}\rangle\langle S_{2}|
\]
\begin{equation}
+J\left(|S_{2}\rangle\langle Q_{x}|+|Q_{x}\rangle\langle S_{2}|\right).\label{eq:QxS2_ham}
\end{equation}
Here, $\epsilon_{Q_{x}}$ and $\epsilon_{S_{2}}$ are the optical
electronic transition energies of the BChl $Q_{x}$ state and the
carotenoid $S_{2}$ state, respectively (including the reorganization
energy of the bath), $H_{Q_{x}}^{(B)}$ and $H_{S_{2}}^{(B)}$ are
the Hamiltonian operators of the nuclear DOF on the BChl and carotenoid,
respectively. These two sets on nuclear DOF described by macroscopic
sets of coordinates $q^{(Q_{x})}=\{q_{1}^{BChl},\dots,q_{n}^{BChl}\text{\}}$
and $q^{(S_{2})}=\{q_{1}^{Car},\dots,q_{n}^{Car}\}$, where $n$ is
a macroscopically large number, form the bath interacting through
operators $\Delta V_{Q_{x}}(q^{(Q_{x})})$ and $\Delta V_{S_{2}}(q^{(S_{2})})$
with the electronic transitions on the BChl and carotenoid, respectively.
The electronic coupling element $J$ describes the resonance interaction
between the collective excited state $|Q_{x}\rangle=|e_{Q_{x}}\rangle|g_{Car}\rangle$
in which the BChl is excited to the state $Q_{x}$ and the carotenoid
is in its electronic ground state and the collective excited state
$|S_{2}\rangle=|g_{BChl}\rangle|e_{S_{2}}\rangle$ in which the BChl
is in its electronic ground state and the carotenoid is excited to
its excited state $S_{2}$. To complete the Hamiltonian we should
also to include the state $|f\rangle=|e_{Q_{x}}\rangle|e_{S_{2}}\rangle$
in which both Car and BChl are excited. However it will be shown later
that the features related to this state are negligible as the resonance
coupling will be shown to be weak. In the experimental 2D spectra,
shown in Fig. \ref{sec:Figure 2D_experimental}, one can easily identify
the fast rising ESA component, which is usually assigned to the strong
absorption from the $S_{1}$ states of the carotenoid. Contribution
of this state is taken into account by including a separate partial
Hamiltonian and $S_{2}\rightarrow S_{1}$ relaxation rate into the
response functions in Appendix \ref{sec:Response-functions} All other
states with no spectroscopic contributions are taken into account
in the simulations by introducing decay rates for the bright states.
For the subsequent discussion it will be useful to introduce a bare
electronic (single exciton) Hamiltonian which reads as
\[
H_{{\rm el}}=\epsilon_{Q_{x}}|Q_{x}\rangle\langle Q_{x}|+\epsilon_{S_{2}}|S_{2}\rangle\langle S_{2}|
\]
\begin{equation}
+J\left(|S_{2}\rangle\langle Q_{x}|+|Q_{x}\rangle\langle S_{2}|\right).\label{eq:Hel}
\end{equation}
This Hamiltonian represents the basic three level (ground state energy
was chosen to be zero) structure of the problem.

\subsection{Energy Transfer Rates}

As discussed above, the interplay of energy gap fluctuations and the
resonance coupling energy determines which theoretical limit applies.
The usual discussion of the relative strengths of the two interactions
is made based on the relative values of the reorganization energy
$\lambda$ which describes the magnitude of the energy gap fluctuations
and the value of the resonance coupling $J$ \cite{Ishizaki2009a}.
Our system is characterized by reorganization energies $\lambda_{Q_{x}}$
of the BChl and $\lambda_{S_{2}}$ of the Car which should be compared
to the resonance coupling energy $J$. For the validity of the weak
resonance coupling limit it is usually required that $\lambda>|J|$.
In our case, assuming $J\approx100$ cm$^{-1}$ (see Ref. \cite{Tretiak2000,Krueger1998})
confirms the validity of this regime because the total $\lambda_{S_{2}}$
is in the order of thousands of cm$^{-1}$ as it involves two fast
vibrational modes with Huang-Rhys factors close to $1$ and vibrational
frequencies above $1000$ cm$^{-1}$ as shown in Fig. \ref{sec:Figure TG}.
The $Q_{y}$ state of BChl in the protein environment of the LH2 complex
exhibits $\lambda_{Q_{y}}\approx100$ cm$^{-1}$ (see e.g. Ref. \cite{Zigmantas2006})
and we can assume a similar value $\lambda_{Q_{x}}\approx100$ cm$^{-1}$
as for the $Q_{x}$ state. Here, the reorganization energy of the
chromophore is only comparable, not larger, than the resonance coupling
element. We note that conventionally the role of temperature is completely
neglected in this discussion. The magnitude of the energy gap correlation
function, which provides a true measure of the energy gap fluctuations,
depends on temperature and the dependency is linear in the high temperature
limit (in the case of Brownian oscillator model, see Ref. \cite{Mukamel1995}).
It would therefore be more suitable to compare the value of $\lambda k_{B}T$
and $|J|^{2}$. At room temperature, $k_{B}T=207$ cm$^{-1}$ and
$\lambda_{Q_{x}}k_{B}T>|J|^{2}$. Yet another problem arises when
considering the role of the relative energy gap $\Delta\varepsilon=\epsilon_{S_{2}}-\epsilon_{Q_{x}}$
of the $Q_{x}$ and $S_{2}$ states. In an ideal dimer, the delocalization
is determined by the so-called mixing angle $\theta=\frac{1}{2}\arctan\frac{2|J|}{|\Delta\epsilon|}.$\cite{May2000}
The delocalization decreases with the increasing $\Delta\varepsilon$.
The interplay of the relative energy gap, temperature, reorganization
energy and resonance energy is a subject of numerous theoretical studies
in the context of photosynthesis. However, no simple formula taking
into account the influences of all these parameters on the degree
of delocalization exists.

After the discussion of the parameters of the carotenoid-BChl dimer,
we have many reasons to believe that the interaction of the $S_{2}$
and $Q_{x}$ states can be treated in the weak coupling limit. The
issues is however more complicated because of the nature of the system-bath
interaction in the carotenoid. Most of its reorganization energy can
be assigned to the fast intramolecular vibrational modes. The spacing
between the vibrational levels and the relative electronic energy
gap $\Delta\varepsilon$ are such that the $Q_{x}$ state can be viewed
as effectively interacting with the nearest vibrational levels and
not with the carotenoid electronic state as a whole. This is a situation
similar to the one studied in Refs. \cite{Christensson2012,Chin2013,Chenu2013}
where special resonances between vibrational and electronic energy
gaps lead to pronounced effects despite the small Huang-Rhys factors
of the BChls. Clearly, treating the carotenoid vibrations as a thermodynamic
bath is a questionable approximation. 

\begin{figure}
\includegraphics[width=1\columnwidth]{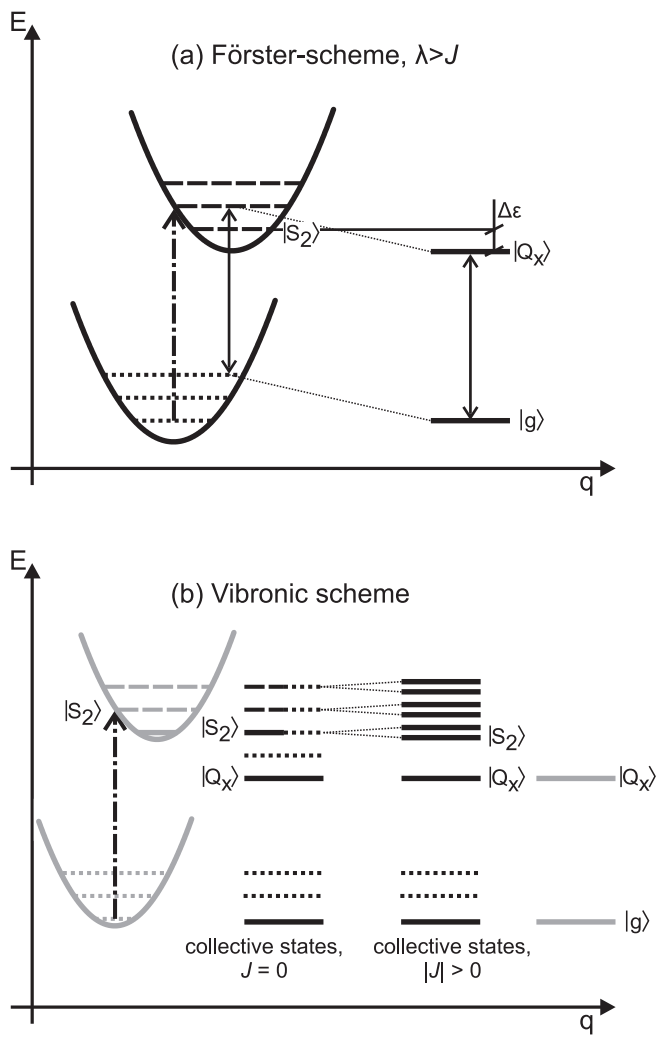}

\protect\caption{\label{fig:Interaction-schemes}Interaction schemes of the electronic
and vibrational levels in (a) localized (F\"{o}rster) and (b) delocalized
(vibronic) regime. In the case (a) the transition energy from the
non-equilibrated state of the carotenoid excited state vibrations
has to be resonant with the $Q_{x}$ transition energy. The relative
energy gap $\Delta\varepsilon$ between the $S_{2}$ and $Q_{x}$
transitions requires involvement of more than one vibrational quanta
of the carotenoid. The vibronic case (b) also involves excitonic mixing
of the vibrational levels of the carotenoid with the $Q_{x}$ transition.
The collective single excited states of the carotenoid-purpurin or
carotenoid-BChl dimer involve states in which the $Q_{x}$ transition
is excited and simultaneously the electronic ground state vibrations
of the carotenoid are excited. The states can excitonically mix with
the carotenoid electronically excited state with a different number
of vibrational quanta. }
\end{figure}

In the following subsections we will briefly discuss the carotenoid-BChl
energy transfer rates based on the weak coupling F\"{o}rster theory, and
compare it with an explicit treatment of the carotenoid vibrations
in the Hamiltonian, along with Refs. \cite{Christensson2012,Chenu2013}.
The energy level structure characteristic for the two situations is
presented in Fig. \ref{fig:Interaction-schemes}.

\subsubsection{Weak Resonance Coupling Limit\label{sub:Weak-Resonance-Coupling}}

In the weak resonance coupling limit, the parameter $J$ of the Hamiltonian,
Eq. (\ref{eq:QxS2_ham}), is assumed small and perturbation theory
to the second order can be performed. For two transitions (donor and
acceptor), the transition rate is given by the Fermi golden rule as$K_{A\leftarrow D}=\frac{2\pi}{\hbar}|J|^{2}\delta(\epsilon_{D}-\epsilon_{A}).$
It is possible to view the donor molecule as a set of many available
de-excitation transition (the molecule is excited electronically and
vibrationally) while the acceptor molecule can be viewed as a set
of transitions ready to be excited. In a realistic case it is therefore
necessary to integrate the Fermi formula over all acceptor and donor
transitions
\[
K_{A\leftarrow D}=\frac{2\pi}{\hbar}|J|^{2}
\]
\begin{equation}
\times\int\limits _{-\infty}^{\infty}{\rm d}\epsilon_{D}\int\limits _{-\infty}^{\infty}{\rm d}\epsilon_{A}f_{{\rm abs}}(\epsilon_{A})f_{{\rm fl}}(\epsilon_{D})\delta(\epsilon_{D}-\epsilon_{A}),\label{eq:KAD_distrib}
\end{equation}
where $f_{{\rm abs}}$ and $f_{{\rm fl}}$ correspond to the normalized
distribution of the transition energies available for excitation on
the acceptor and the normalized distribution of the transition energies
available for the deexcitation on the donor. The F\"{o}rster rate can
be derived directly from Eq. (\ref{eq:KAD_distrib}) by noticing that
the acceptor absorption spectrum is related to the distribution of
transition frequencies $\tilde{f}_{{\rm abs}}(\omega)=f_{{\rm abs}}(\hbar\omega)$
as $\alpha(\omega)\sim\omega\tilde{f}_{{\rm abs}}(\omega)$ and similarly
for the donor fluorescence spectrum $\sigma(\omega)\sim\omega^{3}\tilde{f}_{{\rm fl}}(\omega)$
(see e.g. \cite{VanAmerongen2000,May2000}) or directly by perturbation
theory with respect to $J$ and a cumulant expansion for the bath
DOF as in Ref. \cite{Valkunas2013}. In both cases there is a direct
relation between the rate and the overlap of absorption spectrum of
the acceptor with the emission spectrum of the donor
\begin{equation}
K_{A\leftarrow D}=2\pi\frac{|J|^{2}}{\hbar^{2}}\frac{\int d\omega\alpha_{A}(\omega)\sigma_{D}(\omega)\omega^{-4}}{\int d\omega\alpha_{A}(\omega)\omega^{-1}\int d\omega\sigma_{D}(\omega)\omega^{-3}}.\label{eq:Foerster_final_relaxed}
\end{equation}
It is important to note that the F\"{o}rster theory does not in general
depend on the dipole-dipole coupling approximation although the specific
dependence of the rate in this approximation is often used. From the
point of view of the present manuscript, the coupling energy $J$
can be calculated by any type of advanced methods of quantum chemistry
which takes into account all the details of the electronic structure
of the excited states of the interacting molecules \cite{Scholes2003}.
From the theoretical point of view, the F\"{o}rster rate can be derived
from Hamiltonian, Eq. (\ref{eq:QxS2_ham}), under various assumptions.
Usually one assumes relaxed state of the environment and the intramolecular
vibrational modes. However, F\"{o}rster theory is very versatile and allows
many generalizations \cite{Scholes2003,Beljonne2009}, most importantly
it is still valid for the case that the excited state of the donor
is unrelaxed \cite{Mukamel1995}. Eq. (\ref{eq:KAD_distrib}) is then
still valid, but it does not result in Eq. (\ref{eq:Foerster_final_relaxed}).
F\"{o}rster theory therefore also allows for a direct experimental estimation
of the resonance coupling if the fluorescence spectrum of the donor
and the absorption spectrum of the acceptor are known, even in case
that the fluorescence does not start from a relaxed state. For the
present dyad, the coupling $J$ was determined to be $J=240$ cm$^{-1}$,
within the framework of Eq. (\ref{eq:Foerster_final_relaxed}). \cite{Macpherson2002}

Fig. \ref{fig:Interaction-schemes}(a) describes a typical situation
for the short time dynamics of the $S_{2}-Q_{x}$ energy transfer.
The higher vibrationally excited levels of the $S_{2}$ are populated,
and in order to achieve resonance with the $Q_{x}$ transition, even
higher vibrational levels on the carotenoid ground state are excited
after deexcitation of the donor ($S_{2}$). The energy transfer is
enabled by the fact that the energy corresponding to the $\Delta\varepsilon$
is accepted by the ground state vibrational levels of the carotenoid
molecule as depicted in Fig. \ref{fig:Interaction-schemes}(a). The
same statement can be rephrased in terms of the spectral overlap.
Because the rate depends on the overlap of the donor emission and
the acceptor absorption spectra, the effective broadening of the carotenoid
spectrum due to the fast vibrational modes increases significantly
the energy transfer rate. F\"{o}rster theory thus assigns the electronic
ground state vibrational levels of the carotenoid a crucial importance
in the energy transfer process.

\subsubsection{Vibronic Coupling\label{sub:Vibronic-Coupling}}

The main conclusion of the preceding section is that the vibrational
modes of the carotenoid are the main driving element of the energy
transfer from $S_{2}$ to $Q_{x}$. F\"{o}rster theory shows that these
modes provide resonance for the energy transfer and that this resonance
involves multiple vibrational quanta. As one can learn from Fig. \ref{sec:Figure absorption}(a),
absorption lineshape undergoes a slight change in the carotenoid region
suggesting weak delocalization between $Q_{x}$ and $S_{2}$ states,
at least for dyad. It is therefore natural to include the important
vibrational modes explicitly into the Hamiltonian, and to treat only
the remaining overdamped modes as a bath. This leads directly to the
model of vibronic excitons (see Refs. \cite{Womick2011,Christensson2012,Tiwari2013,Chin2013}).
Such a model automatically treats the selected vibrational modes non-perturbatively. 

We split the nuclear part of the carotenoid Hamiltonian in Eq. (\ref{eq:QxS2_ham})
into some selected modes denoted by coordinates $\tilde{q}_{n}$ and
referred to later in the text as primary modes. The rest of the modes
which we denote by $q^{\prime(S_{2})}$ (see Sec. \ref{sub:Carotenoid-Chlorophyll-Interaction}
for the notation) will be referred to as secondary modes or secondary
bath. We have 
\begin{equation}
H_{S_{2}}^{(B)}=H_{S_{2}}^{(b)}+H_{{\rm vib}}+H_{{\rm vib-b}}.
\end{equation}
where $H_{S_{2}}^{(b)}$ is the Hamiltonian of the secondary bath
and the Hamiltonian of the primary modes and their interaction with
the secondary bath read as
\begin{equation}
H_{{\rm vib}}=\sum_{n}\frac{\hbar\tilde{\omega}_{n}}{2}\left(\tilde{p}_{n}^{2}+\tilde{q}_{n}^{2}\right),
\end{equation}
and 
\begin{equation}
H_{{\rm vib-b}}=-\sum_{n}\tilde{q}_{n}\sum_{k}\xi_{k}^{n}q_{k}.\label{eq:Hvib-b}
\end{equation}
Here, $\tilde{\omega}_{n}$ are the frequencies of the primary modes
to be treated explicitly, and $q_{k}$ are the coordinates from the
set $q^{\prime(S_{2})}$ coupled to the primary modes by some coupling
constants $\xi_{k}^{n}$. In the excited state of the carotenoid,
the bath Hamiltonian is now split in the following way
\[
H_{S_{2}}^{(B)}+\Delta V_{S_{2}}(q^{(S_{2})})=H_{S_{2}}^{(b)}+\Delta V_{S_{2}}(q^{\prime(S_{2})})
\]
\begin{equation}
+\sum_{n}\frac{\hbar\omega_{n}}{2}\left(\tilde{p}_{n}^{2}+(\tilde{q}_{n}-d_{n})^{2}\right)-\sum_{n}(\tilde{q}_{n}-d_{n})\sum_{k}\xi_{k}^{n}q_{k}.
\end{equation}
Here, we assume that the same secondary bath which drives the ground
state vibrations into the equilibrium, drives also the excited state
vibrations to their corresponding equilibrium. Correspondingly, the
energy gap operator reads as
\[
\Delta V_{S_{2}}(q^{(S_{2})})=\Delta V_{S_{2}}(q^{\prime(S_{2})})
\]
\begin{equation}
-\sum_{n}\hbar\tilde{\omega}_{n}d_{n}\tilde{q}_{n}+\sum_{n}d_{n}\sum_{k}\xi_{k}^{(n)}q_{k}.\label{eq:DVS_with_explicit}
\end{equation}
We assumed that the spectral densities describing the interaction
of the selected modes with the rest of the bath are the same. The
energy gap operator, Eq. (\ref{eq:DVS_with_explicit}), is composed
of three parts, $\Delta V_{S_{2}}(q^{\prime(S_{2})})$ representing
the overdamped bath, the part representing the direct influence of
the selected modes on the energy gap fluctuations
\begin{equation}
H_{{\rm el-vib}}=\sum_{n}\hbar\tilde{\omega}_{n}d_{n}\tilde{q}_{n},
\end{equation}
and the part describing the influence of the overdamped bath on the
electronic energy gap functions via the coupling with the selected
modes. The strength of the latter part of the interaction Hamiltonian
is given by the shift $d_{n}$ of the excited state vibrational potential
with respect to the ground state potential. The essence of the vibronic
model is to include $H_{{\rm el-vib}}$ to the Hamiltonian which we
treat explicitly and only an effective energy gap operator
\begin{equation}
\Delta V_{S_{2}}^{({\rm eff)}}(q^{\prime(S_{2})})=\Delta V_{S_{2}}(q^{\prime(S_{2})})+\sum_{n}d_{n}\sum_{k}\xi_{k}^{(n)}q_{k}
\end{equation}
is treated by perturbation theory. It leads to the bath correlation
function 
\begin{equation}
C(t)=\langle\Delta V_{S_{2}}(q^{\prime(S_{2})};t)\Delta V_{S_{2}}(q^{\prime(S_{2})};0)\rangle+C_{{\rm b}}^{({\rm el})}(t),\label{eq:corr_fce_tot}
\end{equation}
where the last term on the right hand side corresponds to the direct
contribution of the secondary bath to the dephasing. The damping of
the $n$th primary modes is governed by the bath correlation function
originating from the term, Eq. (\ref{eq:Hvib-b}), 
\begin{equation}
C_{{\rm b}}^{(n)}(t)=\sum_{k}|\xi_{k}^{n}|^{2}\langle q_{k}(t)q_{k}\rangle,\label{eq:Cbn}
\end{equation}
while the direct contribution of the secondary bath to the dephasing
is expressed by the same correlation function scaled by the dimensionless
shift of the effective vibrational mode
\begin{equation}
C_{{\rm b}}^{({\rm el})}(t)=\sum_{n}|d_{n}|^{2}C_{{\rm b}}^{(n)}(t).\label{eq:Celb}
\end{equation}

The calculation of the energy transfer dynamics in the vibronic model
thus starts with a diagonalization of the Hamiltonian
\begin{equation}
H_{S}=H_{{\rm vib}}+H_{{\rm el-vib}}|S_{2}\rangle\langle S_{2}|+H_{{\rm el}}.
\end{equation}
The diagonalization results in eigenstates which combine the electronic
and vibrational states into the mixed vibronic states. In a second
step the rates of population transfer and coherence dephasing are
calculated for the set of eigenstates by second order perturbation
theory and cumulant expansion \cite{Perlik2014} using correlation
function, Eq. (\ref{eq:Cbn}), for the damping of the selected mode
and the correlation function $C_{{\rm b}}^{({\rm el})}(t)$ only to
simulate electronic dephasing. This means that we assume the correlation
function $C(t)$, Eq. (\ref{eq:corr_fce_tot}), to be composed of
the contribution of the secondary bath only. This choice decreases
the number of degrees of freedom in fitting and effectively ties strength
of the coupling between the primary mode and the secondary bath to
the strengths of the interaction between the electronic transition
and the primary mode. Because only effective spectral density is important
for the calculation of the transition rates, we believe that the advantage
in limiting the number of parameters outweighs the restriction put
on our model. In principle one could estimate the overall rate of
energy transfer directly by weighting the calculated rates between
individual levels. We have however chosen the more practical approach
of calculating the transfer time as described in Section \ref{sub:Calculation-of-Transfer}.

Because of the rather large vibrational frequency and moderate coupling,
the $Q_{x}$ transition can be viewed as interacting only with the
energetically nearest transition. This situation is depicted in Fig.
\ref{fig:Interaction-schemes}(b). For the sake of clarity of the
graphical presentation, we depict a situation of perfect resonance
between the relative energy gap $\Delta\varepsilon$ and the vibrational
frequency and show energy level splitting of the energetically nearest
levels. Fig. \ref{fig:Interaction-schemes}b depicts the relevant
carotenoid and BChl energy levels as two separate entities (in gray)
and in terms of collective states of the carotenoid-BChl dimer in
the case of $J=0$. Conceptually, the ground state vibrational levels
of the carotenoid contribute to an excited collective state if the
$Q_{x}$ state is excited. This is a consequence of the fact in a
singly excited aggregate state all molecules except one are in their
ground state (carotenoid in this case). The ground state vibrational
levels of the carotenoid thus interact resonantly with the vibrational
levels of the carotenoid in the electronically excited state $S_{2}$.
This interaction corresponds to the energy transfer from $S_{2}$
to $Q_{x}$ with a simultaneous deposition of the excess energy to
the ground state vibrations of the carotenoid. The complete picture
of the energy levels of the system is much more complicated because
the excitonic mixing involves all available states. The details of
this mixing, which are included in the basis transformation from the
localized states to the energetic eigenstates, lead to discernible
changes in the lineshape of the carotenoid molecule. As will be demonstrated
in the following sections, the vibronic model provides an explanation
for the observed spectral changes as shown in Fig. \ref{sec:Figure absorption}(a).
Also, the detailed treatment of the interaction between carotenoid
vibrational levels and the $Q_{x}$ state leads to the experimentally
observed energy transfer rates with a significantly weaker resonance
coupling than predicted by Eq. (\ref{eq:Foerster_final_relaxed}).

\subsection{Simulation of Absorption Spectra and Transfer Rates in Vibronic Model}

The vibronic model discussed in the previous section was implemented
as described in Section \ref{sub:Calculation-of-Transfer}. The absorption
spectra of monomeric carotenoid was fitted with a model Hamiltonian
involving one effective high frequency mode to obtain starting values
of the effective mode parameters for the subsequent fitting of the
absorption spectra of the dyad and the LH2 complex. We found the frequency
of the vibrational mode in the dyad to be $\omega_{{\rm eff}}=1390$\,cm$^{-1}$
and its Huang-Rhys factor was determined to be $s_{{\rm eff}}=1.3$.
All monomeric parameters are discussed in Section \ref{sub:Calculation-of-Transfer}.

\begin{figure}[h]
\includegraphics{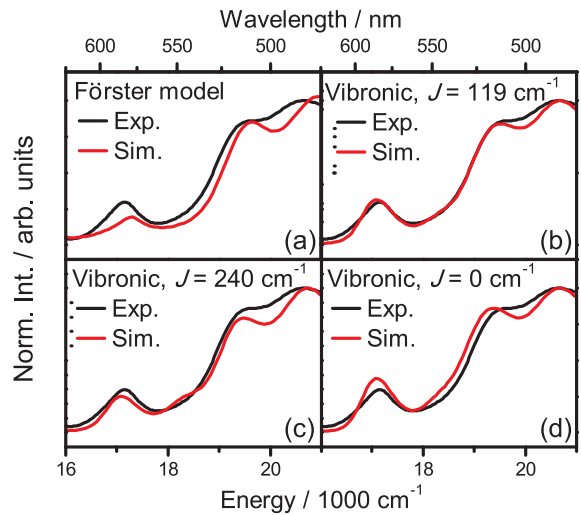}

\protect\caption{\label{fig:Absorption-spectra-dyad}Absorption spectra of the dyad.
Panel (a) shows the sum of the monomeric spectra, which corresponds
to the weak coupling case (no changes to the lineshape except of lifetime
broadening); (b) presents the best fit of the dyad spectrum with the
$Q_{x}$ transition energy, relative energy gap $\Delta\varepsilon$,
carotenoid HR factors and the resonance coupling free to be adjusted.
Panel (c) shows the best fit with the resonance coupling fixed at
$J=240$ cm$^{-1}$ and panel (d) presents the sum of the monomers
with the parameters corresponding to the best fit from panel (b),
but with $J=0$ cm$^{-1}$.}
\end{figure}

In Fig. \ref{fig:Absorption-spectra-dyad} we summarize our fitting
efforts for the dyad. We chose the dyad because the same solvent can
be used for the dyad and its components avoiding solvent related bathochromic
shifts. The absorption spectra of LH2 \emph{M. pur. }and LH2 \emph{Rps.
ac. }were equally well reproduced with the employed vibronic model.
Fig. \ref{fig:Absorption-spectra-dyad}(a) presents the result of
summation of the dyad's monomer spectra. This is the spectrum corresponding
to the weak resonance coupling case (F\"{o}rster regime) and it is presented
to highlight the changes between monomeric and dyad spectra. In Fig.
\ref{fig:Absorption-spectra-dyad}b, we present a result of the fitting
in which the Huang-Rhys factor of the carotenoid, the $Q_{x}$ transition
frequency, the relative energy gap $\Delta\varepsilon$ between $S_{2}$
and $Q_{x}$ and the resonance coupling $J$ were allowed to change.
The fitting achieves good agreement with the experimental dyad spectrum,
most importantly in the carotenoid related part. There the relative
hight of the absorption maxima of different vibrational peaks changes
between the monomer and the dyad. The estimated coupling is $J=-119$\,cm$^{-1}$
while $s_{{\rm eff}}=1.29$, $\Delta\varepsilon=2400$\,cm$^{-1}$
and $\epsilon_{Q_{x}}=17150$\,cm$^{-1}$. The minor change with
respect to the parameters of the monomers, such as the change of the
transition energy to the excited state, have to be assigned to the
chemical change upon formation of the dyad.\cite{Macpherson2002}

The same fitting procedure with the fixed resonance coupling $J$
estimated from measured fluorescence spectrum by F\"{o}rster theory, $J=240$
cm$^{-1}$ \cite{Macpherson2002}, results in a less satisfactory
agreement as can be seen in Fig. \ref{fig:Absorption-spectra-dyad}(c).
In order to demonstrate that lineshape changes in all cases are not
only due to the change of the monomeric Huang-Rhys factor of the carotenoid
molecule, we show the spectrum of the dyad with the parameters resulting
from the same parameters but with $J=0$ cm$^{-1}$ in Fig. \ref{fig:Absorption-spectra-dyad}(d).
Interestingly, the best fit is achieved with a change of the Huang-Rhys
factor of less than $1$ \% and the change in the line shape can be
therefore mostly assigned to the resonance coupling. 

\begin{figure}[h]
\includegraphics{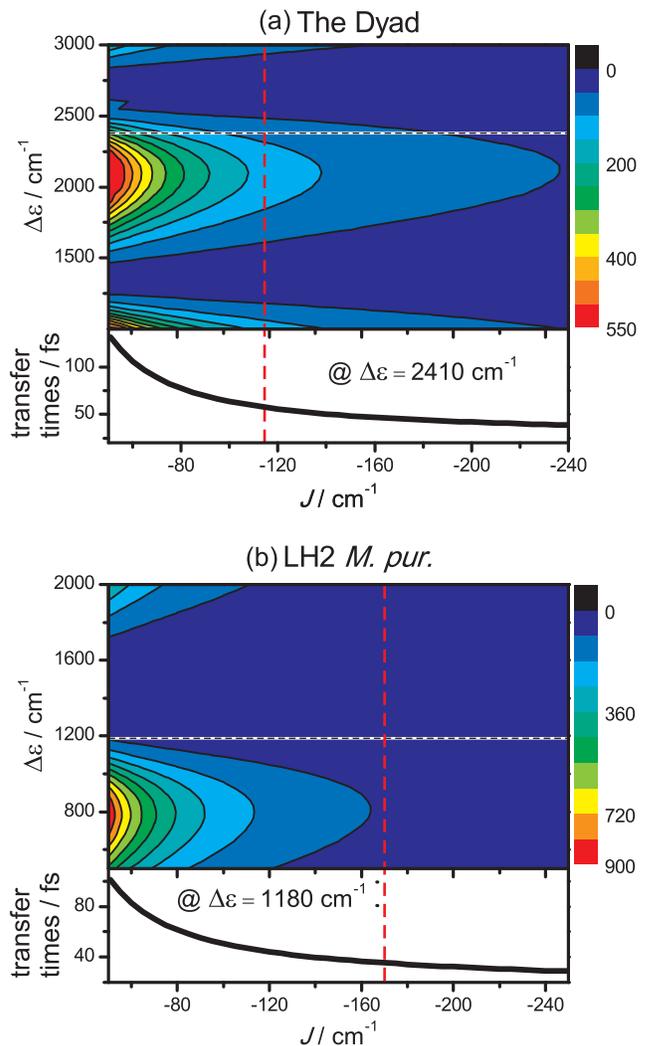}

\protect\caption{\label{fig:Transfer-time}Transfer time as a function of $J$ and
$\Delta\varepsilon$ for the dyad (a) and for LH2 \emph{M. pur.} (b).
At the experimental value of $\Delta\varepsilon$ and the values of
resonance coupling determined by fitting absorption spectrum, transfer
time for the dyad and LH2 \emph{M. pur. }are 55 fs and 40 fs, respectively,
in agreement with the values reported in pump probe measurements. }
\end{figure}

The simulation of the absorption spectrum has resulted in a value
of $J$ which is by $50$ \% smaller than the one estimated from the
experiment, suggesting that higher value is not compatible with the
spectral changes with respect to the monomers seen in the dyad. The
obvious question is whether this lower value for the resonance coupling
is compatible with the measured ultrafast rates. To this end we perform
calculations of the $S_{2}$ to $Q_{x}$ transfer dynamics to estimate
the transfer time introduced in Section \ref{sub:Calculation-of-Transfer}
for various values of $J$ and $\Delta\varepsilon$ and keeping other
parameters of the model fixed to those from the best fit from Fig.
\ref{fig:Absorption-spectra-dyad}(b). The plot of transfer time as
a function of $J$ and $\Delta\varepsilon$ is presented in Fig. \ref{fig:Transfer-time}.
The calculations yield a transfer time of $\tau_{S_{2}\rightarrow Q_{x}}=55$
fs for the values of the best fit which is in good agreement with
the experimental value of $40$ fs.\cite{Macpherson2002} Moreover
the 2D plot in Fig. \ref{fig:Transfer-time} and the corresponding
cut at the value of $J=-119$ cm$^{-1}$reveals a periodic modulation
of the transfer time with the period corresponding to the effective
spacing between vibrational levels. The transfer time decreases monotonically
with increasing value of $J$, as shown in the lower panes of Fig.
\ref{fig:Transfer-time} representing cuts along the experimental
vales of $\Delta\varepsilon$. The periodic modulation of the transfer
time is in agreement with the resonant involvement of the carotenoid
vibrational modes in the energy transfer. As in the F\"{o}rster case this
implies the involvement of the carotenoid vibrational levels from
the electronic ground state. Conservation of energy during the energy
transfer requires the excess energy to be deposited to the vibrational
energy of the carotenoid. 

\subsection{Simulation of the 2D Spectra}

In order to assign the observed features of the experiment 2D-spectrum
we performed simulations of the carotenoid 2D spectra. This is a first
step towards a complete simulation of the interacting Car-BChl system,
which allows us to assign Car only features and possibly identify
2D spectral components originating in $Q_{x}$ state. Such a full
calculation will be a subject of our future work. 

Calculating the 2D spectrum in a spectral region covering the edge
of the studied molecule is an extremely difficult task given the fact
that a good fit on the outer region of (even the absorption) spectrum
depends on many rarely studied details of the chromophore interaction
with its environment. For instance, assumptions about the Gaussian
disorder of the electronic transitions energies and typical estimates
for bath spectral densities are well suited for explaning the features
around the maximum of absorption and do not cover well the situation
at the edges of the spectra. On the other hand, introducing more fitting
freedom by assuming arbitrarily some other types of spectral densities
and disorder distributions seems not to yield more theoretical insight,
as the new parameters cannot be fixed by a limited set of experiments.
Below we therefore describe a calculation of the Car 2D-spectra in
which the Car molecule should well represent the Car component of
the LH2. Correspondingly a limited fitting to the experiment 2D-spectra
from Fig. \ref{sec:Figure 2D_experimental} is done, keeping in mind
that the two systems which are compared in such a fitting are different.
Large part of the Car parameters are therefore motivated by the values
from literature.

\subsubsection{Carotenoid only Spectra}

From the linear absorption spectrum of the LH2 complex the electronic
excitation energy $\omega_{S_2 S_0}=\unit[19770]{cm^{-1}}$ was obtained
(see Fig. \ref{sec:Figure absorption}). Because of the line shape
function approach, which relies on second-order cumulant expansion
with the electronic ground state as the reference state, this value
corresponds to the vertical transition energy, i.e. the sum of the
electronic excitation energy and the reorganization energies of the
spectral density components. The transfer rates were chosen under
the assumptions $1/k_{S_2 \to S_1}=\unit[95]{fs}$ and $1/k_{S_2 \to Q_x}=\unit[55]{fs}$
in agreement with \cite{PoCeLa06_BPJ_2086}. Because of the slow transfer
from $S_1$ to other electronic states, the lifetime broadening constant
$\Gamma_{S_1}$ was taken as zero. Even though some spectral density
parameters for okenone are available from the literature \cite{PoCeLa06_BPJ_2086},
the respective parameters for $\beta$-Carotene have been reported
in more detail. They can at least provide some orientation for the
choice of the parameters to model the okenone component of the investigated
LH2 complex. According to \cite{ChZiMa13_JCP_11209} the frequencies
of the included vibrational modes are $\omega_{UO,1}=\unit[1150]{cm^{-1}}$
and $\omega_{UO,2}=\unit[1520]{cm^{-1}}$. The Huang-Rhys factors
of the vibrational modes in $S_2$ were assumed as $\chi_{UO,S_2 S_2,1}=\unit[0.25]{}$
and $\chi_{UO,S_2 S_2,2}=\unit[0.5]{}$. While these values are somewhat
smaller than the ones for $\beta$-Carotene given in Ref. \cite{ChZiMa13_JCP_11209},
their ratio is similar. According to the tendency reported in the
literature, the Huang-Rhys factors in $S_1$ were assumed to be larger
than those in $S_2$ by a scaling factor chosen as \unit[1.5]{}.
To describe the damping of  the vibrational modes in $S_1$ with a
time constant of $\unit[50]{fs}$, ten Matsubara terms were taken
into account in the calculation of the lineshape function.  For one
of the Brownian oscillator spectral density components a fast decay
with damping constant $1/\Lambda_{BO1,S_2 S_2}=\unit[30]{fs}$ and
a reorganization energy $\lambda_{BO,S_2 S_2,1}=\unit[300]{cm^{-1}}$
in agreement with Ref. \cite{ChMiNe09_JPCB_16409} was assumed. To
reproduce the inhomogeneous broadening effects in the absorption spectrum
of the LH2 complex, the second Brownian oscillator spectral density
component with very slow decay constant of $\unit[20]{ps}$ \cite{ChMiNe09_JPCB_16409}
was included in the model. For a reorganization energy of $\lambda_{BO,S_2 S_2,2}=\unit[3000]{cm^{-1}}$
 comparable inhomogeneous broadening effects as in a measured linear
absorption spectrum could be obtained and features in the measured
2D-spectra could be qualitatively  reproduced. In Ref. \cite{SuYaCo07_PRB_155110}
it has been reported that coherence gets lost during this population
transfer process, so that the fluctuations in $S_2$ and $S_1$ can
be considered as uncorrelated. Accordingly, $g_{S_2 S_1}$ and $g_{S_2 S_n}$
were taken as zero. The Huang-Rhys factors for excitation from $S_1$
to $S_n$  are known to be much smaller than the ones of the excitation
from $S_0$ to $S_2$. In Ref.\cite{PoCeLa06_BPJ_2086} a value of
$\unit[0.1]{}$ is given, which was assumed for both modes in our
calculation. 

According to our interpretation, this value of the Huang-Rhys factor
is given with respect to $S_0$, as the vibrational energy for the
transition between $S_1$ and $S_n$ reported in Ref. \cite{PoCeLa06_BPJ_2086}
cannot be explained by such a small Huang-Rhys factor. However, the
much larger Huang-Rhys factors of $S_1$ with respect to $S_0$ allows
for a larger vibrational energy than expected from the value of $0.1$
for the Huang-Rhys factor.

Regarding the Brownian oscillator modes, it has been reported in \cite{ChMiNe09_JPCB_16409}
that for double excitation from $S_1$ the reorganization energy corresponds
to a fraction of only $\unit[0.5]{}$ of the corresponding reorganization
energy in $S_1$ from the electronic ground state, which was taken
as the same as for $S_2$ in our description.

The vertical transition energy between $S_1$ and $S_n$ was chosen
as $\omega_{S_1 S_n}=\unit[15400]{cm^{-1}}$ in agreement with Ref.
\cite{PoCeLa06_BPJ_2086}. In the prefactors of the ESA response functions
$\left| \frac{\mu_{S_1 S_{n}}}{\mu_{S_0 S_2}} \right|^2=\unit[1.5]{}$
entered, following the values given in \cite{ChPoYaPu09_PRB_245118}
for other carotenoid derivatives. The finite pulse widths were taken
into account in terms of Gaussian profiles determined from a fit of
the measured local oscillator spectrum . In this way the central frequency
$\omega_0=\unit[16275]{cm^{-1}}$ and a FWHM of $\unit[1612]{cm^{-1}}$
were obtained. As the fitted curve corresponds to the squared pulse
profile, multiplication of the FWHM with a factor of $\sqrt{2}$ was
required to obtain the FWHM of the single pulses with a value of $\unit[2280]{cm^{-1}}$.

In the real part of the sum of all calculated response function contributions
after convolution with the pulses an intensive, positive-valued diagonal
peak in the region of ($\omega_1=-\unit[17500]{cm^{-1}}$,$\omega_3=\unit[17500]{cm^{-1}}$)
appears at all population times up to $100$ fs, starting from $t_2=\unit[0]{fs}$.
This peak mainly stems from the rephasing GSB contribution.  Furthermore,
below the diagonal at ($\omega_1=-\unit[17500]{cm^{-1}}$,$\omega_3=\unit[16000]{cm^{-1}}$)a
crosspeak from the rephasing GSB contribution is found, which stems
from vibrational effects. This  initially oval peak   becomes butterfly-shaped
with a nodal line separating positive an negative region at $t_2=\unit[10]{fs}$,
 whereas at $t_2=\unit[20]{fs}$  it becomes completely positive again.
At $t_2=\unit[30]{fs}$ a modified shape of the vibrational crosspeak
below the diagonal  from the rephasing GSB  contribution in combination
with  a rise of of the rephasing SE contribution and the negative-valued
rephasing ESA contribution leads to  a localization of the crosspeak
close to ($\omega_1=-\unit[18000]{cm^{-1}}$,$\omega_3=\unit[15500]{cm^{-1}}$).
 From $t_2=\unit[50]{fs}$ the ESA contribution starts to obscure
the latter.   At $t_2=\unit[80]{fs}$ the negative ESA peak covers
a broad energetic range below the diagonal. This tendency becomes
even more pronounced at $t_2=\unit[100]{fs}$ and $t_2=\unit[200]{fs}$,
where the ESA peak pushes the diagonal peak  almost completely above
the diagonal. 

\begin{figure*}
\includegraphics{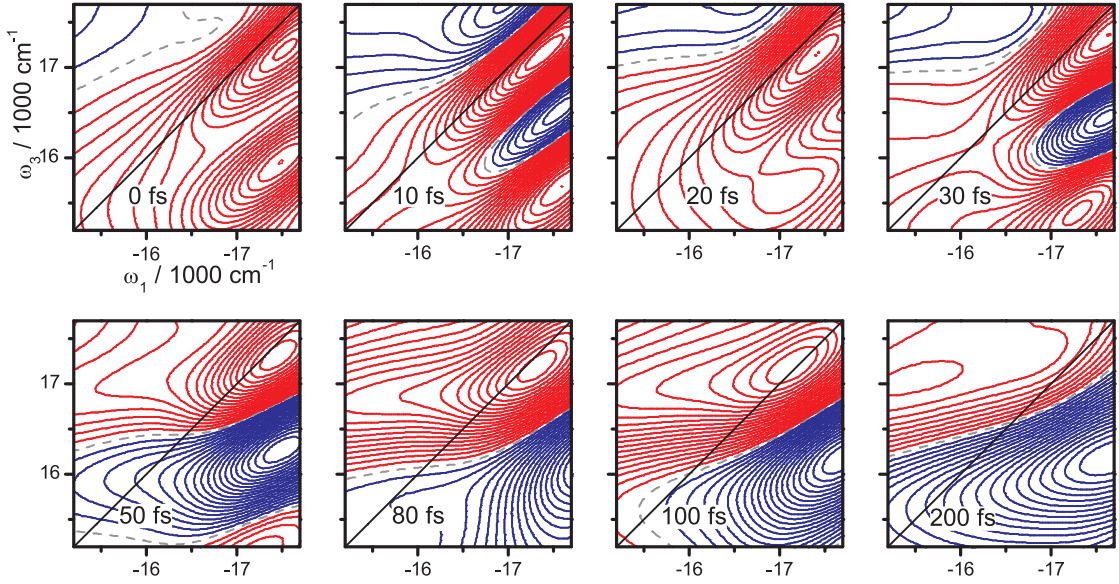}\protect\caption{\label{fig:Simulated-electronic-2D}Simulated electronic 2D spectra
of okenone, the carotenoid in LH2 \textit{M. pur.} The employed values
of $t_{2}$ are indicated in the panels. Line coloring follows the
same conventions as described in Fig.\,\ref{sec:2D-dyad_vs_LH2}}
\end{figure*}

Different from the measured 2D-spectra, the upper diagonal peak at
$t_{2}$\,=\,0 to 30\,fs is missing in the calculated 2D-spectra
shown in Fig.\,\ref{fig:Simulated-electronic-2D}, most likely due
to the fact that the spectral phase of the pulses was considered as
flat in the calculations. When comparing the spectra   from experiment
(Fig.\,\ref{sec:Figure 2D_experimental}) and simulation (Fig.\,\ref{fig:Simulated-electronic-2D})
at $t_{2}$\,=\,30\,fs, it becomes apparent that the experimentally
obtained cross peak, which we identified as an indication of $S_{2}\rightarrow Q_{x}$
energy transfer, is missing in simulations. Considering that the simulations
only incorporate the okenone-part of the spectrum, this missing cross
peak confirms our assignment to energy transfer to $Q_{x}$. 2D-ES
elucidates why carotenoid to bacteriochlorophyll energy transfer pathways
are difficult to analyze in pump-probe spectroscopy. \cite{Macpherson2002,Cong2008}
The according cross peak coincides with a recurring negative carotenoid
feature (see $t_{2}$\,=\,30\,fs and $t_{2}$\,=\,10\,fs spectrum
in Fig.\,\ref{fig:Simulated-electronic-2D}). Additionally, $S_{2}\rightarrow Q_{x}$
energy transfer is overlaid with ESA from $S_{1}$, making it only
observable within the ultrashort lifetime of $S_{2}$, i.e. only during
55\,fs in LH2 \textit{M. pur}.

%%\begin{widetext}

%%\begin{center}

%%\end{center}  %%%\end{widetext}

\section{Conclusions and Outlook}

In this work we have shown that a vibronic coupling mechanism describes
several central aspects of the carotenoid to BChl energy transfer
dynamics, yielding moderate coupling constants $J$ close to 100\,cm$^{-1}$.
This value is in excellent agreement with structure-based calculations
and reproduces experimental absorption spectra well. The low values
of $J$ obtained from the vibronic model suffice for explaining the
experimentally observed ultrafast transfer rates. This is explained
by vibronic resonances, which have the potential of dramatically speeding
up energy transfer. \cite{Womick2011,Butkus2012,Huelga2013} Hence,
the vibronic coupling mechanism circumvents the shortcomings of F\"{o}rster
theory and explains static as well as dynamical properties of natural
light harvesters. 

The investigated carotenoid-BChl system is an extreme case of a heterodimer,
meaning that donor (carotenoid) and acceptor (BChl) differ in transition
energies and HR-factor. Future studies will test the vibronic coupling
mechanism on energy transferring dimers where the spectroscopic properties
are more alike, such as perylene bisimide dyads.\cite{Langhals2010}
In such systems, both donor and acceptor show strong electron-phonon
coupling, which should make vibronic effects in energy transfer dynamics
even more pronounced. Another highly promising class of systems is
bulk heterojunction solar cells, where vibronic coupling between donor
and acceptor was recently suggested to be the mechanism behind ultrafast
electron transfer.\cite{Falke2014}
\begin{acknowledgments}
\appendix
The authors wish to thank Ana Moore for providing the dyad sample
and Heiko Lokstein for fruitful discussion. Funding by the Austrian
Science Fund (FWF): START project Y\,631-N27 (J. H. and C. N. L.
), the CZE - Aus Mobility Grant (Grant No. 7AMB14AT007, CZ 05/2014)
( J. H. and C. N. L. , V. P., J. S., F. S and T.M.) and by the Czech
Science Foundation (GACR) grant. no. 14-25752S (V. P., J. S., F. S
and T.M.) is acknowledged. R. J. C. and L. J. C acknowledge funding
by BBSRC. 
\end{acknowledgments}

\begin{widetext}

\section{Response functions\label{sec:Response-functions}}

In this Appendix, the response functions used to calculated the Car
2D-spectrum are specified. The rephasing GSB term with population
of $S_2$ and a decay due intramolecular transfer to $S_1$ is given
as

\begin{eqnarray} \label{eq:R3_S2_corrected} R_{3g,S_2}(\tau_3,\tau_2,\tau_1) &=& \left| \mu_{S_0 S_2} \right|^4 \exp(i \omega_{S_2 S_0} \tau_1 - i \omega_{S_2 S_0} \tau_3) \exp \left( -\Gamma_{S_2} \tau_1 - \Gamma_{S_2} \tau_3 \right) \nonumber \\ &&\exp(-g^{*}_{S_2 S_2}(\tau_1)+g^{*}_{S_2 S_2}(\tau_2)-g_{S_2 S_2}(\tau_3) \nonumber \\ &&-g^{*}_{S_2 S_2}(\tau_1+\tau_2)-g^{*}_{S_2 S_2}(\tau_2+\tau_3)+g^{*}_{S_2 S_2}(\tau_1+\tau_2+\tau_3)). \end{eqnarray}The
rephasing SE component with decay of the $S_2$ component due to intramolecular
transfer to $S_1$ reads

\begin{eqnarray} \label{eq:R2_S2_decay_corrected} R_{2g,S_2}(\tau_3,\tau_2,\tau_1) &=& \left| \mu_{S_0 S_2} \right|^4 \exp(i \omega_{S_2 S_0} \tau_1 - i \omega_{S_2 S_0} \tau_3) \exp \left( -\Gamma_{S_2} \tau_1 - \Gamma_{S_2} \tau_3 \right) \nonumber \\ &&\exp(-g^{*}_{S_2 S_2}(\tau_1)+g_{S_2 S_2}(\tau_2)-g^{*}_{S_2 S_2}(\tau_3) \nonumber \\ &&-g^{*}_{S_2 S_2}(\tau_1+\tau_2)-g_{S_2 S_2}(\tau_2+\tau_3)+g^{*}_{S_2 S_2}(\tau_1+\tau_2+\tau_3)) \nonumber \\ &&\exp(-k_{S_2 \to S_1} \tau_2). \end{eqnarray}For
the rephasing ESA component with population transfer from $S_2$ to
$S_1$ and subsequent excitation to a higher excited state one obtains 

\begin{eqnarray} \label{eq:R1fcc_S2_S1_corrected} R^{*}_{1f,S_2 \to S_1}(\tau_3,\tau_2,\tau_1) &=& -\left| \mu_{S_0 S_{2}} \right|^2 \left| \mu_{S_1 S_n} \right|^2 \exp(i \omega_{S_2 S_0} \tau_1 - i \omega_{S_n S_1} \tau_3) \exp \left( -\Gamma_{S_2} \tau_1 - \Gamma_{S_1} \tau_3 \right) \nonumber \\ &&\exp(-g^{*}_{S_2 S_2}(\tau_1)-g_{S_2 S_1}(\tau_2)-g_{S_1 S_1}(\tau_3) \nonumber \\ &&+g^{*}_{S_2 S_1}(\tau_1+\tau_2)+g_{S_2 S_1}(\tau_2+\tau_3)-g^{*}_{S_2 S_1}(\tau_1+\tau_2+\tau_3)) \nonumber \\ &&\exp(g_{S_n S_2}(\tau_2)+2g_{S_n S_1}(\tau_3)-g^{*}_{S_n S_2}(\tau_1+\tau_2) \nonumber \\ &&-g_{S_n S_2}(\tau_2+\tau_3)+g^{*}_{S_n S_2}(\tau_1+\tau_2+\tau_3)-g_{S_n S_n}(\tau_3)) \nonumber \\ &&k_{S_2 \to S_1} \int_{0}^{\tau_2} d \tau \exp(-k_{S_2 \to S_1} \tau) \nonumber \\ &&\exp(2i \Im(g_{S_2 S_1}(\tau_2-\tau)-g_{S_1 S_1}(\tau_2-\tau)+g_{S_1 S_1}(\tau_2-\tau+\tau_3)-g_{S_2 S_1}(\tau_2-\tau+\tau_3))) \nonumber \\ &&\exp(2i \Im(g_{S_n S_1}(\tau_2-\tau)-g_{S_n S_2}(\tau_2-\tau)+g_{S_n S_2}(\tau_2-\tau+\tau_3)-g_{S_n S_1}(\tau_2-\tau+\tau_3))). \nonumber \end{eqnarray}The
nonrephasing response functions of GSB-, SE- and ESA-types read

\begin{eqnarray} \label{eq:R4_S2_corrected} R_{4g,S_2}(\tau_3,\tau_2,\tau_1) &=& \left| \mu_{S_0 S_2} \right|^4 \exp(- i \omega_{S_2 S_0} \tau_1 - i \omega_{S_2 S_0} \tau_3) \exp \left( -\Gamma_{S_2} \tau_1 - \Gamma_{S_2} \tau_3 \right) \nonumber \\ &&\exp(-g_{S_2 S_2}(\tau_1)-g_{S_2 S_2}(\tau_2)-g_{S_2 S_2}(\tau_3) \nonumber \\ &&+g_{S_2 S_2}(\tau_1+\tau_2)+g_{S_2 S_2}(\tau_2+\tau_3)-g_{S_2 S_2}(\tau_1+\tau_2+\tau_3)), \end{eqnarray}

\begin{eqnarray} \label{eq:R1_S2_decay_corrected} R_{1g,S_2}(\tau_3,\tau_2,\tau_1) &=& \left| \mu_{S_0 S_2} \right|^4 \exp(- i \omega_{S_2 S_0} \tau_1 - i \omega_{S_2 S_0} \tau_3) \exp \left( -\Gamma_{S_2} \tau_1 - \Gamma_{S_2} \tau_3 \right) \nonumber \\ &&\exp(-g_{S_2 S_2}(\tau_1)-g^{*}_{S_2 S_2}(\tau_2)-g^{*}_{S_2 S_2}(\tau_3) \nonumber \\ &&+g_{S_2 S_2}(\tau_1+\tau_2)+g^{*}_{S_2 S_2}(\tau_2+\tau_3)-g_{S_2 S_2}(\tau_1+\tau_2+\tau_3)) \nonumber \\ &&\exp(-k_{S_2 \to S_1} \tau_2) \end{eqnarray}and

\begin{eqnarray} \label{eq:R2fcc_S2_S1_corrected} R^{*}_{2f,S_2 \to S_1}(\tau_3,\tau_2,\tau_1) &=& -\left| \mu_{S_0 S_{2}} \right|^2 \left| \mu_{S_1 S_n} \right|^2 \exp(- i \omega_{S_2 S_0} \tau_1 - i \omega_{S_n S_1} \tau_3) \exp \left( -\Gamma_{S_2} \tau_1 - \Gamma_{S_1} \tau_3 \right) \nonumber \\ &&\exp(-g_{S_2 S_2}(\tau_1)+g^{*}_{S_2 S_1}(\tau_2)-g_{S_1 S_1}(\tau_3) \nonumber \\ &&-g_{S_2 S_1}(\tau_1+\tau_2)-g^{*}_{S_2 S_1}(\tau_2+\tau_3)+g_{S_2 S_1}(\tau_1+\tau_2+\tau_3)) \nonumber \\ &&\exp(-g^{*}_{S_n S_2}(\tau_2)+2g_{S_n S_1}(\tau_3)+g_{S_n S_2}(\tau_1+\tau_2) \nonumber \\ &&+g^{*}_{S_n S_2}(\tau_2+\tau_3)-g_{S_n S_2}(\tau_1+\tau_2+\tau_3)-g_{S_n S_n}(\tau_3)) \nonumber \\ &&k_{S_2 \to S_1} \int_{0}^{\tau_2} d \tau \exp(-k_{S_2 \to S_1} \tau) \nonumber \\ &&\exp(2i \Im(g_{S_2 S_1}(\tau_2-\tau)-g_{S_1 S_1}(\tau_2-\tau)+g_{S_1 S_1}(\tau_2-\tau+\tau_3)-g_{S_2 S_1}(\tau_2-\tau+\tau_3))) \nonumber \\ &&\exp(2i \Im(g_{S_n S_1}(\tau_2-\tau)-g_{S_n S_2}(\tau_2-\tau)+g_{S_n S_2}(\tau_2-\tau+\tau_3)-g_{S_n S_1}(\tau_2-\tau+\tau_3))). \nonumber \end{eqnarray}

\end{widetext}

\bibliographystyle{prsty}

\end{document}